\documentclass[english,prd,aps,tightenlines,superscriptaddress,amsmath,amssymb,amsfonts,nofootinbib]{revtex4-2}
\usepackage[cp1252]{inputenc}
\usepackage{float}
\usepackage{mathrsfs}
\usepackage{amsmath}
\usepackage{amsthm}
\usepackage{amssymb}
\usepackage{graphicx}
\usepackage{float}

\usepackage{amsfonts}
\usepackage{amsthm}\usepackage{epsfig}\usepackage{color}\usepackage{times}\usepackage[english]{babel}
\usepackage{color}
\usepackage{braket}
\usepackage{tabularx}
\usepackage{array}
\usepackage{xcolor}

\usepackage{babel}
\begin{document}
\title{Quantum Field Theory of neutrino mixing in spacetimes with torsion}

\author{A. Capolupo}
\email{capolupo@sa.infn.it}
\affiliation{Dipartimento di Fisica ``E.R. Caianiello'' Universit\'a degli Studi di Salerno,
  and INFN   Gruppo Collegato di Salerno, Via Giovanni Paolo II, 132, 84084 Fisciano (SA), Italy}

\author{G. De Maria}
\email{giuseppedemaria97@gmail.com}
\affiliation{DIME Sezione Metodi e Modelli Matematici, Universit\`{a} di Genova,
Via All’Opera Pia 15, 16145 Genova, Italy}

\author{S. Monda}
\email{s.monda5@studenti.unisa.it}
\affiliation{Dipartimento di Fisica ``E.R. Caianiello'' Universit\'a degli Studi  di Salerno,
   and INFN  Gruppo Collegato di Salerno, Via Giovanni Paolo II, 132, 84084 Fisciano (SA), Italy}

\author{A. Quaranta}
\email{anquaranta@unisa.it}
\affiliation{Dipartimento di Fisica ``E.R. Caianiello'' Universit\'a degli Studi  di Salerno, 
  and INFN   Gruppo Collegato di Salerno, Via Giovanni Paolo II, 132, 84084 Fisciano (SA), Italy}

\author{R. Serao}
\email{rserao@unisa.it}
\affiliation{Dipartimento di Fisica ``E.R. Caianiello'' Universit\'a degli Studi  di Salerno, 
  and INFN   Gruppo Collegato di Salerno, Via Giovanni Paolo II, 132, 84084 Fisciano (SA), Italy}

\begin{abstract}
In the framework of quantum field theory, we analyze the neutrino
oscillations in the presence of a torsion background. We consider
the Einstein-Cartan theory and we study the cases of constant torsion and of linearly time dependent torsion. We derive
new neutrino oscillation formulae which are depending on the spin
orientation. Indeed the energy splitting induced by the torsion influences
oscillation amplitudes and frequencies. This effect is maximal for
values of torsion of the same order of the neutrino masses and for
very low momenta, and disappears for large values of torsion. Moreover,
neutrino oscillation is inhibited for intensities of torsion term
much larger than neutrino masses and momentum. The modifications induced
by torsion on the $CP$-asymmetry has been also presented. Future
experiments, such as PTOLEMY, could provide insights into the effect
shown here.
\end{abstract}

\maketitle 

\section{Introduction}

Theories of gravity beyond General Relativity (GR) have a long and
complex history \cite{Capoz1}. Stimulated by the need of dealing
with the shortcomings of GR, providing an explanation for the dark
components of the universe, and possibly to set a viable framework
for the quantization of gravity, there is by now a plethora of such
theories. Some, as the early attempt to incorporate Mach's principle
by Brans and Dicke \cite{BransDicke}, involve additional fields other
than the metric \cite{ST1,ST2}. Other theories generalize the Einstein-Hilbert
action, eventually including higher order curvature invariants \cite{HO1}.
Quite a natural generalization of GR emerges when one considers a
non symmetric connection, allowing for the possibility of torsion
\cite{Hehl,Shapiro}. Gravitational theories including torsion might
be able to account for dark matter and dark energy \cite{DETorsion}.
Torsion couples naturally to the spin density of matter, inducing
a spin-dependent splitting of the energy levels \cite{Cabral} and
spin oscillations \cite{Cirillo-Lombardo}.

Neutrinos, on the other hand, have a prominent role in cosmology and
astrophysics. Their comparatively small interaction rates and the
abundance in which they are produced make neutrinos a precious source
of information on the cosmos. They are possibly linked to the original
baryon asimmetry \cite{Lepto1}, to dark matter \cite{Curv0,Curv0-1} and
dark energy \cite{Kaplan}. Neutrinos also pose several challenges
to the standard model of particles, and many aspects of neutrino physics,
including the basic mechanism behind flavor oscillations \cite{N1,N2,N3,N4,N5-0,N5,N5-1,N6,N7,N8,N9,N10,N11,N12,N13,N14,N15,N16,N17},
the origin of their mass and their fundamental nature \cite{NBSM1,NBSM2},
are yet to be clarified.

In this paper we analyze the propagation of neutrinos on a torsion
background and study its impact on the flavor oscillations. Neutrino
oscillations in presence of torsion have been studied in the quantum
mechanical framework \cite{NTorsion1,NTorsion2}. We here approach
the subject from the point of view of quantum field theory and quantise
the neutrino fields on a torsion background. We focus on the simplest
generalization of GR including torsion, the Einstein-Cartan theory.
We consider the cases of constant torsion and of torsion linearly depending on time, and we assume that spacetime
curvature is absent. We show that the energy splitting induced by
the torsion term leads to spin-dependent neutrino oscillation formulae.
Indeed, the spin orientation affects the frequencies, as expected
also in QM framework, and the oscillation amplitudes which in QFT
are ruled by the Bogoliubov coefficients. This last effect is a pure
consequence of the non-trivial condensate structure induced by neutrino
mixing in QFT.

The spin dependence of the oscillation formulae is maximal for intensities
of torsion comparable to the neutrino masses.  On the
other hand, much larger values of torsion carry out to flavor oscillations
which are identical for the two spins, since they become essentially
independent of the spin. Another effect is that a torsion large enough
can effectively inhibit the flavor oscillations, since in this case
the energy differences due to the various masses become irrelevant
with respect to the common torsional energy term. The presence of
torsion is more relevant on neutrino oscillations in non-relativistic
regimes, for which the QFT effects are also more emphasized. Some
phenomenological consequences of the theoretical results presented
here could then be provided, in the future, by experiments that analyze
non-relativistic neutrinos, such as PTOLEMY \cite{N17,Ptolemy}. We
additionally discuss the modifications induced by torsion on the $CP$-asymmetry,
which is a byproduct of the Dirac $CP$-violating phase in the mixing
matrix. We show that also the $CP$ asymmetry depends on the spin
orientations in presence of the torsion background.

The paper is structured as follows. In section II we introduce the
concept of spacetime torsion and we quantize a Dirac field on torsional
background. In section III, we analyze the three flavor neutrino mixing
in the presence of constant and time dependent torsion, and in section IV, we derive new
oscillation formulae depending
on the orientation of the spin and in section V, new expressions of $CP$ violation are shown. The last section is devoted to the
conclusions, while in the appendix we report the analysis of currents
and charges for flavor mixing in the presence of torsion.

\section{Spacetime Torsion and Dirac field quantization}

Here, we briefly recall the notion of spacetime torsion, then we
quantize the Dirac field minimally coupled to the torsion in the framework of the Einstein-Cartan theory. We study the cases of constant and time-dependent torsion.

\subsection{Spacetime Torsion}

In general relativity, the requirements of metricity of the covariant
derivative and of symmetry uniquely determine the connection coefficients
(Christoffel symbols) in terms of the metric as follows:
\[
\Gamma_{\mu\nu}^{\rho}=\frac{1}{2}g^{\rho\sigma}\left(\partial_{\mu}g_{\sigma\nu}+\partial_{\nu}g_{\sigma\mu}-\partial_{\sigma}g_{\mu\nu}\right)=\Gamma_{\nu\mu}^{\rho}\ .
\]
A more general theory, the so called Einstein-Cartan (or Riemann-Cartan
geometry), is obtained if the assumption of symmetry is relaxed, keeping
only metricity. In this case, the connection coefficients acquire
an antysimmetric part given by
\begin{eqnarray}
\tilde{\Gamma}_{\mu\nu}^{\rho}-\tilde{\Gamma}_{\nu\mu}^{\rho}=T_{\mu\nu}^{\rho}\ \ \ \ \ \ ;\ \ \ \ \ \ \tilde{\Gamma}_{\mu\nu}^{\rho}=\Gamma_{\mu\nu}^{\rho}+K_{\mu\nu}^{\rho}\ ,
\end{eqnarray}
where the tensors $T_{\mu\nu}^{\rho}$ and $K_{\rho\mu\nu}=\frac{1}{2}\left(T_{\rho\mu\nu}+T_{\mu\nu\rho}-T_{\nu\rho\mu}\right)$
are respectively known as torsion and contorsion. It is also convenient
to introduce \cite{Shapiro} the trace vector $V_{\mu}=T_{\mu\rho}^{\rho}$,
the axial vector $T^{\mu}=\epsilon^{\alpha\beta\gamma\mu}T_{\alpha\beta\gamma}$
and the tensor $q_{\mu\nu}^{\rho}$, in terms of which the torsion
is expressed as
\[
T_{\rho\mu\nu}=\frac{1}{3}\left(V_{\mu}g_{\rho\nu}-V_{\nu}g_{\rho\mu}\right)-\frac{1}{6}\epsilon_{\rho\mu\nu\sigma}T^{\sigma}+q_{\rho\mu\nu}\ ,
\]
and the scalar curvature reads as
\[
\tilde{R}=R-2\nabla_{\mu}V^{\mu}-\frac{4}{3}V_{\mu}V^{\mu}+\frac{1}{2}q_{\rho\mu\nu}q^{\rho\mu\nu}+\frac{1}{24}T_{\mu}T^{\mu}\ .
\]
Here $R$, is the general relativistic Ricci scalar given in terms
of the metric. Notice that the covariant derivatives in this context
are the usual ones involving only the Christoffel symbols. The vacuum
action for Einstein-Cartan is given by the natural generalization
of the Einstein-Hilbert action. It is written as
\begin{equation}
S_{EC}=-\frac{1}{\kappa^{2}}\int d^{4}x\sqrt{-g}\tilde{R}\ ,\label{ECAction}
\end{equation}
with $\kappa=\frac{8\pi G}{c^{4}}$. The torsion-related terms in
Eq. \eqref{ECAction} form a total derivative, not contributing to
the field equations. As a consequence the vacuum theory is equivalent
to general relativity. On the other hand, the situation changes in
presence of matter, where a coupling of the form
\begin{equation}
S_{Tm}=\int d^{4}x\sqrt{-g}K_{\mu\nu}^{\rho}\Sigma_{\rho}^{\mu\nu}
\end{equation}
appears. The spin tensor, here denoted with $\Sigma_{\rho}^{\mu\nu}$,
is constructed out of matter fields. We point out that, the field
equations obtained by varying the total action with respect to contorsion
simply lead to the algebraic constraint $K_{\rho\mu\nu}\propto\Sigma_{\rho\mu\nu}$,
expressing the proportionality of torsion and spin angular momentum.
In the following we will be interested in Dirac spinors minimally
coupled to torsion. The spin covariant derivatives, in presence of
torsion, get modified as follows \cite{Cabral}
\begin{equation}
\tilde{D}_{\mu}\psi=D_{\mu}\psi+\frac{1}{4}K_{AB\mu}\left[\gamma^{A},\gamma^{B}\right]\psi
\end{equation}
where $D_{\mu}$ is the general relativistic spin covariant derivative
and the Lorentz indices on the contorsion tensor result from contraction
with the tetrads $K_{AB\mu}=e_{A}^{\rho}e_{B}^{\sigma}K_{\rho\sigma\mu}$.
Then, the spinor action is simply given by
\begin{equation}
\tilde{S}_{D}=S_{D}+S_{TD}=\int d^{4}\sqrt{-g}\left[\frac{i}{2}\left(\bar{\psi}\gamma^{\mu}D_{\mu}\psi-D_{\mu}\bar{\psi}\gamma^{\mu}\psi\right)-m\bar{\psi}\psi\right]+3\int d^{4}x\sqrt{-g}T_{\mu}S^{\mu}\label{TotalDiracAction}
\end{equation}
where $S_{D}$ is the Dirac action in general relativity and $S_{TD}=3\int d^{4}x\sqrt{-g}T_{\mu}S^{\mu}$
is the action term due to the Dirac - torsion coupling. Moreover,
$S^{\mu}=\frac{1}{2}\bar{\psi}\gamma^{\mu}\gamma^{5}\psi$ is the
Dirac spin vector. We remark that in all the above expressions the
spacetime dependence of the curved gamma matrices is kept implicit
$\gamma^{\mu}=\gamma^{\mu}(x)=e_{A}^{\mu}(x)\gamma^{A}$.

\subsection{Dirac field quantization on constant torsional background}

From now on we shall assume that some astrophysical source other than
the Dirac field itself generates a background torsion. As far as minimally
coupled Dirac fields are concerned, the information about torsion
is stored in the axial vector field $T^{\mu}(x)$. Since we are specifically
interested in the effects of torsion on Dirac fields, we will assume
that spacetime curvature is absent (although the most general case
can be treated in a similar fashion, see e.g. \cite{Curv0,Curv1,Curv2,Curv3,Curv4,Curv5}),
so that the covariant derivatives in \eqref{TotalDiracAction} are
replaced with standard derivatives and the gamma matrices reduce to
the flat ones. Under these assumptions the Dirac equation becomes
\begin{eqnarray}
i\gamma^{\mu}\partial_{\mu}\psi & = & m\psi-\frac{3}{2}T_{\rho}\gamma^{\rho}\gamma^{5}\psi\ .\label{DiracEquation}
\end{eqnarray}
Canonical quantization proceeds as in flat spacetime, and the Dirac
field may be expanded on any complete set of solutions of Eq. \eqref{DiracEquation}.
We shall see that the expansion closely resembles that of flat spacetime
when a constant torsion background is considered. It is important
to remark that the lepton charge $Q=\int d^{3}x\bar{\psi}\gamma^{0}\psi$
is conserved as a consequence of the $U(1)$ gauge invariance of the
action \eqref{TotalDiracAction}.

In this subsection, we deal with the simplest possible torsion background.
We consider a constant axial torsion directed along the third spatial
axis. The study of time dependent torsion background will be carried out below. The Dirac equation for constant torsion reads
\begin{eqnarray}
i\gamma^{\mu}\partial_{\mu}\psi & = & m\psi-\frac{3}{2}T_{3}\gamma^{3}\gamma^{5}\psi\ ,\label{DiracEquation1}
\end{eqnarray}
and is solved \cite{Cabral} in momentum space by the spinors
\begin{equation}
u_{\vec{k}}^{\uparrow}=N^{+}\left(\begin{array}{c}
1\\
0\\
\frac{k_{3}}{E_{\vec{k}}^{+}+\widetilde{m}^{+}}\\
\frac{k_{1}+ik_{2}}{E_{\vec{k}}^{+}+\widetilde{m}^{+}}
\end{array}\right)\qquad u_{\vec{k}}^{\downarrow}=N^{-}\left(\begin{array}{c}
0\\
1\\
\frac{k_{1}-ik_{2}}{E_{\vec{k}}^{-}+\widetilde{m}^{-}}\\
-\frac{k_{3}}{E_{\vec{k}}^{-}+\widetilde{m}^{-}}
\end{array}\right)\qquad v_{\vec{k}}^{\uparrow}=N^{+}\left(\begin{array}{c}
\frac{k_{3}}{E_{\vec{k}}^{+}+\widetilde{m}^{+}}\\
\frac{k_{1}+ik_{2}}{E_{\vec{k}}^{+}+\widetilde{m}^{+}}\\
1\\
0
\end{array}\right)\qquad v_{\vec{k}}^{\downarrow}=N^{-}\left(\begin{array}{c}
\frac{k_{1}-ik_{2}}{E_{\vec{k}}^{-}+\widetilde{m}^{-}}\\
-\frac{k_{3}}{E_{\vec{k}}^{-}+\widetilde{m}^{-}}\\
0\\
1
\end{array}\right)\ .\label{MSolutions}
\end{equation}
These solutions are formally the same as in flat space, except for
a spin-dependent mass term $\widetilde{m}^{\pm}=m\pm\frac{3}{2}T^{3}$.
The torsion has indeed the effect of lifting the degeneracy in energy
between the two spin orientations $E_{\vec{k}}^{\pm}=\sqrt{\vec{k}^{2}+\widetilde{m}^{\pm^{2}}}$.
By fixing the normalization to $u_{\vec{k}}^{r\dagger}u_{\vec{k}}^{r}=1=v_{\vec{k}}^{r\dagger}v_{\vec{k}}^{r}$,
the factors $N^{\pm}$ are determined as $N^{\pm}=\sqrt{\frac{E^{\pm}+\widetilde{m}^{\pm}}{2E^{\pm}}}$.
Setting $u_{\vec{k}}^{r}(t)=e^{-iE^{r}t}u_{\vec{k}}^{r}$ and $v_{\vec{k}}^{r}(t)=e^{iE^{r}t}v_{\vec{k}}^{r}$,
the Dirac field is expanded as
\begin{equation}
\psi(\vec{x},t)=\sum_{r}\int\frac{d^{3}k}{(2\pi)^{\frac{3}{2}}}\left(u_{\vec{k}}^{r}(t)\alpha_{\vec{k}}^{r}+v_{-\vec{k}}^{r}(t)\beta_{-\vec{k}}^{r\dagger}\right)e^{i\vec{k}\cdot\vec{x}}\label{FreeFieldExpansion}
\end{equation}
with the coefficients obeying the canonical anticommutation relations.
Since the solutions to eq.(\ref{DiracEquation}) are similar to those
obtained in flat space time, to derive the neutrino oscillation formulas
in the presence of torsion, we can follow a procedure analogous to
the one presented in ref.\cite{N5} where the oscillation formulas
for neutrinos in quantum field theory in flat space were found.
Here, we obtain new oscillation formulae,
showing a behavior different with respect to the ones of ref.\cite{N5}. The differences are all contained in the Bogoliubov coefficients
which characterize the amplitudes of the oscillation formulae and which are depending on the spin orientation.

\subsection{Dirac field quantization with time-dependent torsion}

We now quantize the Dirac field coupled to a certain class of time-dependent of torsional backgrounds, namely with $\breve{T}^0$ spacetime constant and the spatial components $\breve{T}^{i} (t) \ , \ \ i=1,2,3$ having an arbitrary time dependency (yet retaining constancy with respect to the spatial variables). This class of backgrounds allows for a simple non-trivial generalization of the constant torsion treatment presented above. For concreteness we treat in some more detail the case of a linearly time-dependent torsion, i.e. $\breve{T}^{i} = \alpha^{i}t$ for some constants $\alpha^{i}$. The Dirac equation is formally equivalent to \eqref{DiracEquation}
\[
\left(i\gamma^{\mu}\partial_{\mu}-m\right)\Psi(x)=\eta\breve{T}^{\rho} (t) \gamma_{\rho}\gamma^{5}\Psi(x)\ ,
\]
except for the explicit dependency of the torsion on time. 
In order to derive the solution of the Dirac equation with torsion,
we write the spinor in the following form
\[
\Psi(x)=\sum_{\lambda}\int d^{3}p\left(A_{\vec{p},\lambda}u_{\vec{p},\lambda}(t,\vec{x})+B_{-\vec{p},\lambda}^{\dagger}v_{\vec{p},\lambda}(t, \vec{x})\right)\,.
\]
\\
We use the ansatz
$u_{\vec{p},\lambda}(t,\boldsymbol{x})=e^{i\boldsymbol{p}\cdot\boldsymbol{x}}\left(\begin{array}{c}
f_{p}(t)\xi_{\lambda}(\hat{p})\\
g_{p}(t)\lambda\xi_{\lambda}(\hat{p})
\end{array}\right)$, for positive energy and $v_{\vec{p},\lambda}(t,\boldsymbol{x})=e^{i\boldsymbol{p}\cdot\boldsymbol{x}}\left(\begin{array}{c}
g_{p}^{*}(t)\xi_{\lambda}(\hat{p})\\
-f_{p}^{*}(t)\lambda\xi_{\lambda}(\hat{p})
\end{array}\right)$ for negative energy. Here $\xi_{\lambda} (\hat{p})$ denote the helicity eigenspinors, satisfying $\left(\vec{\sigma} \cdot \hat{p}\right) \xi_{\lambda}(\hat{p}) = \lambda \xi_{\lambda}(\hat{p})$ for $\lambda = \pm$.  Then, the solution of the Dirac equation  is determined by solving the following system:
\begin{equation}
i\partial_{t}\left(\begin{array}{c}
f_{\vec{p}}(t)\\
g_{\vec{p}}(t)
\end{array}\right)=\left(\begin{array}{cc}
m-\eta\lambda\breve{T}^{i}\hat{p}^{i} & p+\eta\lambda\breve{T}^{0}\\
p+\eta\lambda\breve{T}^{0} & -m-\eta\lambda\breve{T}^{i}\hat{p}^{i}
\end{array}\right)\left(\begin{array}{c}
f_{\vec{p}}(t)\\
g_{\vec{p}}(t)
\end{array}\right)\equiv\mathscr{A}(t)\left(\begin{array}{c}
f_{\vec{p}}(t)\\
g_{\vec{p}}(t)
\end{array}\right)\label{eq: sistema per f e g}
\end{equation}
\\
The eigenvalues of the matrix in eq.(\ref{eq: sistema per f e g})  are $h_{1,2}=\eta\lambda\breve{T}^{i}\hat{p}^{i}\pm\sqrt{m^{2}+\left(p+\eta\lambda\breve{T}^{0}\right)^{2}}$
and the eigenvectors are $v_{1}^{\lambda}=C_{1}\left(\begin{array}{c}
\frac{m+\sqrt{m^{2}+\left(p+\eta\lambda\breve{T}^{0}\right)^{2}}}{p+\eta\lambda\breve{T}^{0}}\\
1
\end{array}\right)$ and $v_{2}^{\lambda}=C_{2}\left(\begin{array}{c}
\frac{m-\sqrt{m^{2}+\left(p+\eta\lambda\breve{T}^{0}\right)^{2}}}{p+\eta\lambda\breve{T}^{0}}\\
1
\end{array}\right)$, with normalization relations: $\left(v_{1}^{\lambda}\right)^{\dagger}v_{1}^{\lambda}=1$
e $\left(v_{2}^{\lambda}\right)^{\dagger}v_{2}^{\lambda}=1$. \\
If   $[\mathscr{A}(t), \mathscr{A}(t') ]= 0$ for $t\neq t'$, then the system of eqs.$(\ref{eq: sistema per f e g})$
can be solved by means of a simple exponentiation:
\begin{equation}
\left(\begin{array}{c}
f_{\vec{p}}(t)\\
g_{\vec{p}}(t)
\end{array}\right)=\exp\left\{ -i\int_{0}^{t}\mathscr{A}(\tau)d\tau\right\} \left(\begin{array}{c}
f_{\vec{p}}(0)\\
g_{\vec{p}}(0)
\end{array}\right)\,.\label{eq: soluzioni per f e g}
\end{equation}
It is here that the requirement of constancy of $\breve{T}^0$ becomes relevant, since the condition $\left[\mathscr{A}(t),\mathscr{A}(t')\right]=0$
is fulfilled for
$\breve{T}^{0}$  independent of time (i.e. $\breve{T}^{0}=\alpha^{0}$). 
The solutions can be explicitly written as
\begin{equation}
\left\{ \begin{array}{c}
f_{\vec{p},\lambda}(t)=\exp\left\{ -i\eta\lambda \hat{p}^i\int_0^t d\tau\breve{T}^i(\tau)\right\} \exp\left\{ -i\omega_{p,\lambda}t\right\} C_{\vec{p},\lambda}\\
g_{\vec{p},\lambda}(t)=\frac{p+\eta\lambda\breve{T}^{0}}{\left(\omega_{p,\lambda}+m\right)}\exp\left\{ -i\eta\lambda \hat{p}^i\int_0^t d\tau\breve{T}^i(\tau)\right\} \exp\left\{ -i\omega_{p,\lambda}t\right\} C_{\vec{p},\lambda} \ ,
\end{array}\right.\label{eq: soluzioni f e g, generale}
\end{equation}
for some constant $C_{\pmb{p},\lambda}$ and $\omega_{\pmb{p},\lambda} = \sqrt{m^2 + \left(p + \eta \lambda \breve{T}^0 \right)^2}$. 
In the specific case of $\breve{T}^{i} = \alpha^i t$ one has
\begin{equation}
\left\{ \begin{array}{c}
f_{\vec{p},\lambda}(t)=\exp\left\{ -i\frac{t^{2}}{2}\eta\lambda\alpha^{i}\hat{p}^{i}\right\} \exp\left\{ -i\omega_{p,\lambda}t\right\} C_{\vec{p},\lambda}\\
g_{\vec{p},\lambda}(t)=\frac{p+\eta\lambda\breve{T}^{0}}{\left(\omega_{p,\lambda}+m\right)}\exp\left\{ -i\frac{t^{2}}{2}\eta\lambda\alpha^{i}\hat{p}^{i}\right\} \exp\left\{ -i\omega_{p,\lambda}t\right\} C_{\vec{p},\lambda}
\end{array}\right.\label{eq: soluzioni f e g}
\end{equation}
By imposing the normalisation condition $\left|f_{\vec{p},\lambda}(t)\right|^{2}+\left|g_{\vec{p},\lambda}(t)\right|^{2}=\frac{1}{\left(2\pi\right)^{3}}$,
we determine $C_{\vec{p},\lambda}=\frac{\omega_{p,\lambda}+m}{\left(2\pi\right)^{\frac{3}{2}}\sqrt{\left(\omega_{p,\lambda}+m\right)^{2}+
\left(p+\eta\lambda\breve{T}^{0}\right)^{2}}}\,.$\\

\section{FLAVOR MIXING WITH TORSION}

In this section, we  analyze the three-flavor neutrino mixing in the presence of torsion, in particular we consider the cases of constant and time dependent torsion. In both the two cases, the neutrino fields
with definite masses $\Psi_{m}^{T}\equiv(\nu_{1},\nu_{2},\nu_{3})$ satisfy the equation
\begin{equation}
i\gamma^{\mu}\partial_{\mu}\Psi_{m}-M_{d}\Psi_{m}=-\frac{3}{2}{T}^{3}\gamma_{3}\gamma^{5}\Psi_{m}\ ,\label{mt}
\end{equation}
with $M_{d}\equiv diag(m_{1},m_{2},m_{3})$. The fields with definite
masses shall be expanded as in eq.\eqref{FreeFieldExpansion}, except
for acquiring an additional label $j=1,2,3$ distinguishing the mass
($u_{\vec{k},j}^{r},\alpha_{\vec{k},j}^{r},...$). The flavor fields
are obtained by performing the appropriate $SU(3)$ rotation on the
mass triplet. We choose the CKM parametrization of the PNMS matrix
to link the the triplet of flavor fields $\psi_{f}^{T}=\left(\nu_{e},\nu_{\mu},\nu_{\tau}\right)$
to the fields with definite masses $\Psi_{m}^{T}$. As shown in ref.\cite{N5}
the rotation to flavor fields can be recast in terms of the mixing
generator $\emph{I}_{\theta}$ as $\nu_{\sigma}^{\alpha}=\emph{I}_{\theta}^{-1}(t)\nu_{i}^{\alpha}(x)\emph{I}_{\theta}(t)\,,$
where $(\sigma,i)=(e,1),(\mu,2),(\tau,3)$, and $\emph{I}_{\theta}(t)=\emph{I}_{23}(t)\emph{I}_{13}(t)\emph{I}_{12}(t)\;.$
For reader convenience, we report in the appendix A the explicit form of the formulae used in the computations.

 We note that the generator $\emph{I}_{\theta}^{-1}(t)$ here introduced,
is formally identical to the generator $G_{\theta}^{-1}(t)$ presented
in ref.\cite{N5}, where the mixing of three families of neutrinos
in flat space-time has been studied. The difference consists in the
fact that while $G_{\theta}^{-1}(t)$ of ref.\cite{N5} is expressed
in terms of the Dirac fields in flat space-time, $\emph{I}_{\theta}^{-1}(t)$
contains Dirac fields which are the solution of the Dirac equations for fields in the presence of torsions (constant and time depentent).
As we will see below, this leads to two new set of Bogoliubov coefficients, one for constant torsion and one for time depending torsio, which
  are dependent on the spin. At the operational
level, $\emph{I}_{\theta}^{-1}(t)$ shares the same properties as
$G_{\theta}^{-1}(t)$. However, it is essential to underline that,
despite the formal analogy, the result here obtained presents completely
new behaviors, since the new neutrino oscillation formulas, which
will be derived in the following, have amplitudes and frequencies
depending on the spin orientation. This effect, due to the torsion, is also depending on the form of the torsion and
can in principle affect neutrinos produced in the nuclei of spiral
galaxies or in rotating black holes.

In the following, adopting the procedure used in ref.\cite{N5}, and
taking into account the presence of torsion, we show the intermediate
steps to derive the new oscillation formulae  and we show the different behaviors of the oscillation formulae for  the adopted torsions. We start by recalling some properties of the mixing generator $\emph{I}_{\theta}^{-1}(t)$
shared with $G_{\theta}^{-1}(t)$.  $\emph{I}_{\theta}^{-1}(t)$ is a map between
the Hilbert space of free fields $\mathcal{H}_{1,2,3}$ and that of
interacting fields $\mathcal{H}_{e,\mu,\tau}$: $\emph{I}_{\theta}^{-1}(t)\,:\,\mathcal{H}_{1,2,3}\rightarrow\mathcal{H}_{e,\mu,\tau}\,.$
At finite volume, the vacuum $\left|0\right\rangle _{1,2,3}$, relative
to the space $\mathcal{H}_{1,2,3}$, is connected to the vacuum $\left|0\right\rangle _{e,\mu,\tau}$,
relative to the space $\mathcal{H}_{e,\mu,\tau}$, in the following
way: $\left|0(t)\right\rangle _{e,\mu,\tau}=\emph{I}_{\theta}^{-1}(t)\left|0\right\rangle _{1,2,3}\,,$
where $\left|0\right\rangle _{e,\mu,\tau}$ is the vacuum for the flavour fields. The explicit form of $\emph{I}_{\theta}^{-1}(t)$ is reported in the appendix A.  The action of the mixing generator defines the plane wave
expansion of the flavor fields
\[
\nu_{\sigma}(x)=\sum_{r}\int\frac{d^{3}\boldsymbol{k}}{(2\pi)^{\frac{3}{2}}}\left[u_{\vec{k},i}^{r}\alpha_{\vec{k},\nu_{\sigma}}^{r}(t)+v_{-\vec{k},i}^{r}\beta_{-\vec{k},\nu_{\sigma}}^{r\dagger}(t)\right]\exp\{i\vec{k}\cdot\vec{x}\}\qquad\qquad\sigma=1,2,3
\]
where the flavor annihilators are given by $\alpha_{\vec{k},\nu_{\sigma}}^{r}(t)\equiv\emph{I}_{\theta}^{-1}(t)\alpha_{\vec{k},i}^{r}\emph{I}_{\theta}(t)\,,
\quad\beta_{-\vec{k},\nu_{\sigma}}^{r\dagger}(t)\equiv\emph{I}_{\theta}^{-1}(t)\beta_{-\vec{k},i}^{r\dagger}(t)\emph{I}_{\theta}(t)\,.$
By definition, they annihilate the flavor vacuum $\alpha_{\vec{k},\nu_{\sigma}}^{r}\left|0\right\rangle _{f}=0=\beta_{-\vec{k},\nu_{\sigma}}^{r}\left|0\right\rangle _{f}$
and, being the above transformations canonical, they satisfy the equal
time canonical anticommutation relations.
The explicit relations of the the flavor annihilators, for $\vec{k}=(0,0,\left|\vec{k}\right|)$, are:
\begin{align*}
\alpha_{\vec{k},\nu_{e}}^{r}(t) & =c_{12}c_{13}\alpha_{\vec{k},1}^{r}+s_{12}c_{13}\left(\left(\Gamma_{12;\vec{k}}^{rr}(t)\right)^{*}\alpha_{\vec{k},2}^{r}+
\varepsilon^{r}\left(\Sigma_{12;\vec{k}}^{rr}(t)\right)\beta_{-\vec{k},2}^{r\dagger}\right)\\
 & +e^{-i\delta}s_{13}\left(\left(\Gamma_{13;\vec{k}}^{rr}(t)\right)^{*}\alpha_{\vec{k},3}^{r}+\varepsilon^{r}\left(\Sigma_{13;\vec{k}}^{rr}(t)\right)
 \beta_{-\vec{k},3}^{r\dagger}\right)\,,
 \\
\alpha_{\vec{k},\nu_{\mu}}^{r}(t) & =\left(c_{12}c_{23}-e^{i\delta}s_{12}s_{23}s_{13}\right)\alpha_{\vec{k},2}-\left(s_{12}c_{23}+e^{i\delta}c_{12}s_{23}s_{13}\right)\times\\
 & \times\left(\left(\Gamma_{12;\vec{k}}^{rr}(t)\right)\alpha_{\vec{k},1}^{r}-\varepsilon^{r}\left(\Sigma_{12;\vec{k}}^{rr}(t)\right)
 \beta_{-\vec{k},1}^{r\dagger}\right)+s_{23}c_{13}\left(\left(\Gamma_{23;\vec{k}}^{rr}(t)\right)^{*}\alpha_{\vec{k},3}^{r}+\varepsilon^{r}
 \left(\Sigma_{23;\vec{k}}^{rr}(t)\right)\beta_{-\vec{k},3}^{r\dagger}\right)\,,
 \\
\alpha_{\vec{k},\nu_{\tau}}^{r}(t) & =c_{23}c_{13}\alpha_{\vec{k},3}^{r}-\left(c_{12}s_{23}+e^{i\delta}s_{12}c_{23}s_{13}\right)
\left(\left(\Gamma_{23;\vec{k}}^{rr}(t)\right)\alpha_{\vec{k},2}^{r}-
\varepsilon^{r}\left(\Sigma_{23;\vec{k}}^{rr}(t)\right)\beta_{-\vec{k},2}^{r\dagger}\right)+\\
 & +\left(s_{12}s_{23}-e^{i\delta}c_{12}c_{23}s_{13}\right)
 \left(\left(\Gamma_{13;\vec{k}}^{rr}(t)\right)\alpha_{\vec{k},1}^{r}-\varepsilon^{r}\left(\Sigma_{13;\vec{k}}^{rr}(t)\right)
 \beta_{-\vec{k},1}^{r\dagger}\right)\,,\\
\\
\beta_{-\vec{k},\nu_{e}}^{r}(t) & =c_{12}c_{13}\beta_{-\vec{k},1}(t)+s_{12}c_{13}\left(\left(\Gamma_{12;\vec{k}}^{rr}(t)\right)^{*}\beta_{-\vec{k},2}^{r}-
\varepsilon^{r}\left(\Sigma_{12;\vec{k}}^{rr}(t)\right)\alpha_{\vec{k},2}^{r\dagger}\right)+\\
 & +e^{i\delta}s_{13}\left(\left(\Gamma_{13;\vec{k}}^{rr}(t)\right)^{*}\beta_{-\vec{k},3}^{r}-\varepsilon^{r}
 \left(\Sigma_{13;\vec{k}}^{rr}(t)\right)\alpha_{\vec{k},3}^{r\dagger}\right)\,,\\
\beta_{-\vec{k},\nu_{\mu}}^{r}(t) & =\left(c_{12}c_{23}-e^{-i\delta}s_{12}s_{23}s_{13}\right)\beta_{-\vec{k},2}^{r}-\left(s_{12}c_{23}+e^{-i\delta}c_{12}s_{23}s_{13}\right)\times\\
 & \times\left(\left(\Gamma_{12;\vec{k}}^{rr}(t)\right)\beta_{-\vec{k},1}^{r}+\varepsilon^{r}
 \left(\Sigma_{12;\vec{k}}^{rr}(t)\right)\alpha_{\vec{k},1}^{r\dagger}\right)+s_{23}c_{13}
 \left(\left(\Gamma_{23;\vec{k}}^{rr}(t)\right)^{*}\beta_{-\vec{k},3}^{r}-\varepsilon^{r}\left(\Sigma_{23;\vec{k}}^{rr}(t)\right)
 \alpha_{\vec{k},3}^{r\dagger}\right)\,,
 \\
\beta_{-\vec{k},\nu_{\tau}}^{r}(t) & =c_{23}c_{13}\beta_{-\vec{k},3}^{r}-\left(c_{12}s_{23}+e^{-i\delta}s_{12}c_{23}s_{13}\right)\left(\left(\Gamma_{23;\vec{k}}^{rr}(t)\right)
\beta_{-\vec{k},2}^{r}+\varepsilon^{r}\left(\Sigma_{23;\vec{k}}^{rr}(t)\right)\alpha_{\vec{k},2}^{r\dagger}\right)+\\
 & +\left(s_{12}s_{23}-e^{-i\delta}c_{12}c_{23}s_{13}\right)\left(\left(\Gamma_{13;\vec{k}}^{rr}(t)\right)\beta_{-\vec{k},1}^{r}+
 \varepsilon^{r}\left(\Sigma_{13;\vec{k}}^{rr}(t)\right)\alpha_{\vec{k},1}^{r\dagger}\right)\,.
\end{align*}
\\
The Bogoliubov coefficients $ \Gamma_{ij;\vec{k}}^{rr}$ and $ \Sigma_{ij;\vec{k}}^{rr}$, appearing in the expressions of  the
flavor annihilators, are given by the inner product of the solutions of Dirac equations with different masses. In order to distinguish
 the case of constant torsion from that of time-dependent torsion, we use the following notation:
   $ \Gamma_{ij;\vec{k}}^{rr} = \Xi_{ij;\vec{k}}^{rr} $ and
 $ \Sigma_{ij;\vec{k}}^{rr} = \chi_{ij;\vec{k}}^{rr} $,
 for constant torsion, and $ \Gamma_{ij;\vec{k}}^{rr} = \Pi_{ij;\vec{k}}^{rr} $ and $ \Sigma_{ij;\vec{k}}^{rr} = \Upsilon_{ij;\vec{k}}^{rr} $, for time dependent torsion. The explicit form of the Bogoliubov coefficients in the two cases analyzed are reported in the following.

 \subsection{Bogoliubov coefficients with constant torsion}

For constant torsion, the modules of the  Bogoliubov coefficients are given by
\[
\left|\Xi_{i,j;\vec{k}}^{r,s}\right|\equiv u_{\vec{k},i}^{r\text{\ensuremath{\dagger}}}u_{\vec{k},j}^{s}=v_{-\vec{k},i}^{s\text{\ensuremath{\dagger}}}v_{-\vec{k},j}^{r}\,,
\qquad\left|\chi_{i,j;\vec{k}}^{r,s}\right|\equiv\varepsilon^{r}u_{\vec{k},1}^{r\text{\ensuremath{\dagger}}}
v_{-\vec{k},2}^{s}=-\varepsilon^{r}u_{\vec{k},2}^{r\text{\ensuremath{\dagger}}}v_{-\vec{k},1}^{s}\,.
\]
Notice that, in
the reference frame $\vec{k}=(0,0,\left|\vec{k}\right|)$, $\Xi_{i,j;\vec{k}}^{r,s}$ and $\chi_{i,j;\vec{k}}^{r,s}$ vanish for $r\neq s$. Explicitly one has:
\[
\begin{array}{c}
\Xi_{ij;\vec{k}}^{\pm\pm}=N_{i}^{\pm}N_{j}^{\pm}\left[1+\frac{k^{2}}{\left(E_{\vec{k},i}^{\pm}+\widetilde{m}_{i}^{\pm}\right)
\left(E_{\vec{k},j}^{\pm}+\widetilde{m}_{j}^{\pm}\right)}\right]=\cos(\xi_{ij;\vec{k}}^{\pm \pm})\,,\\
\chi_{ij;\vec{k}}^{\pm\pm}=N_{i}^{\pm}N_{j}^{+}\left[\frac{k_{3}}{E_{\vec{k},j}^{\pm}+\widetilde{m}_{j}^{\pm}}-\frac{k_{3}}{E_{\vec{k},i}^{\pm}+
\widetilde{m}_{i}^{\pm}}\right]=\sin(\xi_{ij;\vec{k}}^{\pm\pm})\,,
\end{array}
\]
\\
 with the spin-dependent masses and the normalisation coefficients
given explicitly by $\widetilde{m}_{i}^{\pm}\equiv m_{i}\pm\frac{3}{2}T^{3}$
and $N_{i}^{\pm}=\frac{\sqrt{E_{\vec{k},i}^{\pm}+\widetilde{m}_{i}^{\pm}}}{\sqrt{2E_{\vec{k},i}^{\pm}}}$,
respectively. The sign factor is defined as $\varepsilon^{\pm}=\mp1$.
Additionally, $(E_{\vec{k},i}^{\pm})^{2}=\vec{k}^{2}+(\widetilde{m}_{i}^{\pm})^{2}$
and $\xi_{ij;\vec{k}}^{\pm\pm}=\arctan\left(\frac{\left|V_{ij;\vec{k}}^{\pm\pm}\right|}{\left|U_{ij;\vec{k}}^{\pm\pm}\right|}\right)$.
  The canonicity of the Bogoliubov transformations
 is ensured by the relations
$
\sum_{r}\left(\left|\Xi_{ij;\vec{k}}^{\pm r}\right|^{2}+\left|\chi_{ij;\vec{k}}^{\pm r}\right|^{2}\right)=1
$
where $i,j=1,2,3$ and $j>i$. Moreover, the time dependence of $\Xi_{ij;\vec{k}}^{\pm r}$ and $\chi_{ij;\vec{k}}^{\pm r}$  is expressed by
\[
\Xi_{ij;\vec{k}}^{rs}(t)=\left|\Xi_{ij;\vec{k}}^{rs}\right|e^{i\left(E_{\vec{k},j}^{s}-
E_{\vec{k},i}^{r}\right)t}\;,\:\;\;\;\;\chi_{ij;\vec{k}}^{rs}(t)=\left|\chi_{ij;\vec{k}}^{rs}\right|e^{i\left(E_{\vec{k},j}^{s}+
E_{\vec{k},i}^{r}\right)t}\,.
\]
\\

\subsection{Bogoliubov coefficients with time dependent torsion}

In this case, the Bogoliubov coefficients are denoted with $\varPi_{ij;\vec{k}}^{rs}(t)=\left(u_{\vec{k},i}^{r},u_{\vec{k},j}^{s}\right)_{t}$
and $\varUpsilon_{ij;\vec{k}}^{rs}(t)=\left(u_{\vec{k},i}^{r},v_{\vec{k},j}^{s}\right)_{t}$.
The mixed coefficients are zero and explicitly we have:
\begin{align}
\varPi_{ij;\vec{p}}^{++}(t) & =\left(2\pi\right)^{3}\exp\left\{ -i\left(\omega_{p,+}^{j}-\omega_{p,+}^{i}\right)t\right\} \left(C_{\vec{p},i}^{+}\right)^{*}\left(C_{\vec{p},j}^{+}\right)\left[1+\frac{\left|p+\eta\breve{T}^{0}\right|^{2}}{\left(\omega_{p,+}^{i}+m_{i}\right)\left(\omega_{p,+}^{j}+m_{j}\right)}\right]\,,\label{eq: Bogoljubov U++ time-dependent}\\
\varPi_{ij;\vec{p}}^{--}(t) & =\left(2\pi\right)^{3}\exp\left\{ -i\left(\omega_{p,-}^{j}-\omega_{p,-}^{i}\right)t\right\} \left(C_{\vec{p},i}^{-}\right)^{*}\left(C_{\vec{p},j}^{-}\right)\left[1+\frac{\left|p-\eta\breve{T}^{0}\right|^{2}}{\left(\omega_{p,-}^{i}+m_{i}\right)\left(\omega_{p,-}^{j}+m_{j}\right)}\right]\,,\nonumber \\
\varUpsilon_{ij;\vec{p}}^{++}(t) & =\left(2\pi\right)^{3}\exp\left\{ +it^{2}\eta\alpha^{i}\hat{p}^{i}\right\} \exp\left\{ +i\left(\omega_{p,+}^{j}+\omega_{p,+}^{i}\right)t\right\} \left(C_{\vec{p},i}^{+}\right)^{*}\left(C_{\vec{p},j}^{+}\right)^{*}\left(p+\eta\breve{T}^{0}\right)\left[\frac{1}{\omega_{p,+}^{j}+m_{j}}-\frac{1}{\omega_{p,+}^{i}+m_{i}}\right]\,,\label{eq: Bogoljubov V++ time-dependent}\\
\varUpsilon_{ij;\vec{p}}^{--}(t) & =\left(2\pi\right)^{3}\exp\left\{ +it^{2}\eta\alpha^{i}\hat{p}^{i}\right\} \exp\left\{ +i\left(\omega_{p,+}^{j}+\omega_{p,+}^{i}\right)t\right\} \left(C_{\vec{p},i}^{+}\right)^{*}\left(C_{\vec{p},j}^{+}\right)^{*}\left(p-\eta\breve{T}^{0}\right)\left[\frac{1}{\omega_{p,+}^{j}+m_{j}}-\frac{1}{\omega_{p,+}^{i}+m_{i}}\right]\,,\nonumber
\end{align}
\\
where $i,j=1,2,3$ and $j>i$.\footnote{ 
In the ultrarelativistic case ($p \gg m_j$), one has:
\[
\varPi_{\vec{p}}^{rr}(t)\longrightarrow1\,,\;\hfill\varUpsilon_{\vec{p}}^{rr}(t)\longrightarrow0\,
\]
for any $t$. Moreover, in the absence of torsion (i.e. $\breve{T}^{\mu}=0$)
these coefficients coincide  with those presented in the Minkowski metric.}
The canonicity of the Bogoliubov transformations
is satisfied by the following relations
$
\sum_{r}\left(\left|\varPi_{ij;\vec{k}}^{\pm r}\right|^{2}+\left|\varUpsilon_{ij;\vec{k}}^{\pm r}\right|^{2}\right)=1\,.
$

\section{NEUTRINO OSCILLATIONS WITH BACKGROUND TORSION}

In this section, we derive the neutrino oscillation formulae in the presence of torsion and  we study, in particular, the two cases of constant and linear time dependent torsion.
By analyzing flavor currents and charges in a way similar  to what was done in the  ref.\cite{N5}, and as shown in appendix A,  we can define the flavor charges in the presence of torsion as
$
::\,Q_{\nu_{\sigma}}\,::=\sum_{r}\int d^{3}\boldsymbol{k}\left(\alpha_{\vec{k},\nu_{\sigma}}^{r\dagger}(t)\alpha_{\vec{k},\nu_{\sigma}}^{r}(t)-
\beta_{\vec{k},\nu_{\sigma}}^{r\dagger}(t)\beta_{\vec{k},\nu_{\sigma}}^{r}(t)\right)\,,$ with $\sigma=e,\mu,\tau
$
 and, $\left.::\cdots::\right.$, denoting the normal ordering with respect
to the flavor vacuum state $\left|0\right\rangle _{f}$.

The oscillation formulas are obtained by computing, in the Heisenberg picture, the expectation values
of the above charges on the (flavor) neutrino state, defined at $t=0$, as $\left|\nu_{\vec{k},\sigma}^{r\dagger}(0)\right\rangle =\alpha_{\vec{k},\nu_\sigma}^{r\dagger}(0)\left|0\right\rangle _{f}$. At a fixed momentum $\vec{k}$ they are given by:
$
\mathcal{Q}_{\nu_{\rho}\rightarrow\nu_{\sigma}}^{r,\vec{k}}(t)  \equiv\left\langle \nu_{\vec{k},\rho}^{r}(t)\right|::\,Q_{\nu_{\sigma}}\,::\left|\nu_{\vec{k},\rho}^{r}(t)\right\rangle -{}_{f}\left\langle 0\right|::\,Q_{\nu_{\sigma}}\,::\left|0\right\rangle _{f}\,,
$ and explicitly
\begin{align}
\mathcal{Q}_{\nu_{e}\rightarrow\nu_{e}}^{r,\vec{k}}(t) & =1-\sin^{2}(2\theta_{12})\cos^{4}(\theta_{13})\left[\left|\Gamma_{12;\vec{k}}^{rr}\right|^{2}\sin^{2}\left(\Delta_{12;\vec{k}}^{r}t\right)
+\left|\Sigma_{12;\vec{k}}^{rr}\right|^{2}\sin^{2}\left(\Omega_{12;\vec{k}}^{r}t\right)\right]\nonumber \\
 & -\sin^{2}(2\theta_{13})\cos^{2}(\theta_{12})\left[\left|\Gamma_{13;\vec{k}}^{rr}\right|^{2}\sin^{2}\left(\Delta_{13;\vec{k}}^{r}t\right)
 +\left|\Sigma_{13;\vec{k}}^{rr}\right|^{2}\sin^{2}\left(\Omega_{13;\vec{k}}^{r}t\right)\right]\nonumber \\
 & -\sin^{2}(2\theta_{13})\sin^{2}(\theta_{12})\left[\left|\Gamma_{23;\vec{k}}^{rr}\right|^{2}\sin^{2}\left(\Delta_{23;\vec{k}}^{r}t\right)
 +\left|\Sigma_{23;\vec{k}}^{rr}\right|^{2}\sin^{2}\left(\Omega_{23;\vec{k}}^{r}t\right)\right]\,,\label{eq: Formula di oscillazione 3 generazioni Pe-e}
\end{align}
\begin{align}
\mathcal{Q}_{\nu_{e}\rightarrow\nu_{\mu}}^{r,\vec{k}}(t) & =2J_{CP}\left[\left|\Gamma_{12;\vec{k}}^{rr}\right|^{2}\sin\left(2\Delta_{12;\vec{k}}^{r}t\right)-
\left|\Sigma_{12;\vec{k}}^{rr}\right|^{2}\sin\left(2\Omega_{12;\vec{k}}^{r}t\right)\right.
+\left(\left|\Gamma_{12;\vec{k}}^{rr}\right|^{2}-\left|\Sigma_{13;\vec{k}}^{rr}\right|^{2}\right)\sin\left(2\Delta_{23;\vec{k}}^{r}t\right)\nonumber \\
 &+
 \left(\left|\Sigma_{12;\vec{k}}^{rr}\right|^{2}-\left|\Sigma_{13;\vec{k}}^{rr}\right|^{2}\right)\sin\left(2\Omega_{23;\vec{k}}^{r}t\right)
 \left.- \left|\Gamma_{13;\vec{k}}^{rr}\right|^{2}\sin\left(2\Delta_{13;\vec{k}}^{r}t\right)+\left|\Sigma_{13;\vec{k}}^{rr}
 \right|^{2}\sin\left(2\Omega_{13;\vec{k}}^{r}t\right)\right]
 \nonumber \\
  & +\cos^{2}(\theta_{13})\sin(\theta_{13})\left[\cos\delta\sin(2\theta_{12})\sin(2\theta_{23})+
 4\cos^{2}(\theta_{12})\sin\theta_{13}\sin^{2}\theta_{23}\right]\times\nonumber \\
 & \times\left[\left|\Gamma_{13;\vec{k}}^{rr}\right|^{2}\sin^{2}(\Delta_{13;\vec{k}}^{r}t)+\left|\Sigma_{13;\vec{k}}^{rr}\right|^{2}\sin^{2}
 (\Omega_{13;\vec{k}}^{r}t)\right]\nonumber \\
 & -\cos^{2}\theta_{13}\sin\theta_{13}\left[\cos\delta\sin(2\theta_{12})\sin(2\theta_{23})-
 4\sin^{4}\theta_{12}\sin\theta_{13}\sin^{2}\theta_{23}\right]\times\nonumber \\
 & \times\left[\left|\Gamma_{23;\vec{k}}^{rr}\right|^{2}\sin^{2}(\Delta_{23;\vec{k}}^{r}t)+
 \left|\Sigma_{23;\vec{k}}^{rr}\right|^{2}\sin^{2}(\Omega_{23;\vec{k}}^{r}t)\right]\nonumber \\
 & +\cos^{2}\theta_{13}\sin(2\theta_{12})\left[\left(\cos^{2}\theta_{23}-\sin^{2}\theta_{23}\sin^{2}\theta_{13}\right)\sin(2\theta_{12})
 \right.\nonumber \\
 & \left.+\cos\delta\cos(2\theta_{12})\sin\theta_{13}\sin(2\theta_{23})\right]\left[\left|\Gamma_{12;\vec{k}}^{rr}\right|^{2}\sin^{2}
 (\Delta_{12;\vec{k}}^{r}t)+\left|\Gamma_{12;\vec{k}}^{rr}\right|^{2}\sin^{2}(\Omega_{12;\vec{k}}^{r}t)\right]\,,\label{eq: Formula di oscillazione 3 generazioni Pe-mu}
\end{align}
\begin{align}
\mathcal{Q}_{\nu_{e}\rightarrow\nu_{\tau}}^{r,\vec{k}}(t)=- & 2J_{CP}\left[\left|\Gamma_{12;\vec{k}}^{rr}\right|^{2}\sin\left(2\Delta_{12;\vec{k}}^{r}t\right)-
\left|\Sigma_{12;\vec{k}}^{rr}\right|^{2}\sin\left(2\Omega_{12;\vec{k}}^{r}t\right)\right.
+\left(\left|\Gamma_{12;\vec{k}}^{rr}\right|^{2}-\left|\Sigma_{13;\vec{k}}^{rr}\right|^{2}\right)\sin\left(2\Delta_{23;\vec{k}}^{r}t\right)\nonumber \\
 &
 +\left(\left|\Sigma_{12;\vec{k}}^{rr}\right|^{2}-\left|\Sigma_{13;\vec{k}}^{rr}\right|^{2}\right)\sin\left(2\Omega_{23;\vec{k}}^{r}t\right)
 \left.-\left|\Gamma_{13;\vec{k}}^{rr}\right|^{2}\sin\left(2\Delta_{13;\vec{k}}^{r}t\right)+\left|\Sigma_{13;\vec{k}}^{rr}\right|^{2}
 \sin\left(2\Omega_{13;\vec{k}}^{r}t\right)\right]\nonumber \\
 & -\cos^{2}(\theta_{13})\sin(\theta_{13})\left[\cos\delta\sin(2\theta_{12})\sin(2\theta_{23})-
 4\cos^{2}(\theta_{12})\sin\theta_{13}\cos^{2}\theta_{23}\right]\times\nonumber \\
 & \times\left[\left|\Gamma_{13;\vec{k}}^{rr}\right|^{2}\sin^{2}(\Delta_{13;\vec{k}}^{r}t)+\left|\Sigma_{13;\vec{k}}^{rr}\right|^{2}
 \sin^{2}(\Omega_{13;\vec{k}}^{r}t)\right]\nonumber \\
 & +\cos^{2}\theta_{13}\sin\theta_{13}\left[\cos\delta\sin(2\theta_{12})\sin(2\theta_{23})+
 4\sin^{4}\theta_{12}\sin\theta_{13}\cos^{2}\theta_{23}\right]\times\nonumber \\
 & \times\left[\left|\Gamma_{23;\vec{k}}^{rr}\right|^{2}\sin^{2}(\Delta_{23;\vec{k}}^{r}t)+
 \left|\Sigma_{23;\vec{k}}^{rr}\right|^{2}\sin^{2}(\Omega_{23;\vec{k}}^{r}t)\right]\nonumber \\
 & +\cos^{2}\theta_{13}\sin(2\theta_{12})\left[\left(\sin^{2}\theta_{23}-\sin^{2}\theta_{13}\cos^{2}\theta_{23}\right)
 \sin(2\theta_{12})\right.\nonumber \\
 & \left.-\cos\delta\cos(2\theta_{12})\sin\theta_{13}\sin(2\theta_{23})\right]
 \left[\left|\Gamma_{12;\vec{k}}^{rr}\right|^{2}\sin^{2}(\Delta_{12;\vec{k}}^{r}t)+
 \left|\Sigma_{12;\vec{k}}^{rr}\right|^{2}\sin^{2}(\Omega_{12;\vec{k}}^{r}t)\right]\,.\label{eq: Formula di oscillazione 3 generazioni Pe-tau}
\end{align}
where $r= \pm$, $\Delta_{ij;\vec{k}}^{r}\equiv\frac{E_{j;\vec{k}}^{r}-E_{i;\vec{k}}^{r}}{2}\;,$ $\Omega_{ij;\vec{k}}^{r}\equiv\frac{E_{j;\vec{k}}^{r}+E_{i;\vec{k}}^{r}}{2}\,,$ and $J_{CP}$ denotes the Jarlskog invariant $J_{CP}\equiv\mathrm{Im}\left(u_{i\alpha}u_{j\beta}u_{i\beta}^{*}u_{j\alpha}^{*}\right)\,.$
In the parameterization  under consideration, $J_{CP}$ is given by:
$
J_{CP}=\frac{1}{8}\sin\delta\sin(2\theta_{12})\sin(2\theta_{13})\cos(\theta_{13})\sin(2\theta_{23})\,.
$
Notice that, $\mathcal{Q}_{\nu_{\rho}\rightarrow\nu_{e}}^{r,\vec{k}}(t)
+\mathcal{Q}_{\nu_{\rho}\rightarrow\nu_{\mu}}^{r,\vec{k}}(t)
+\mathcal{Q}_{\nu_{\rho}\rightarrow\nu_{\tau}}^{r,\vec{k}}(t)=1\,.$
It is also easy to check that the above oscillation formulae reduce to the Pontecorvo formulae
in absence of torsion in the ultrarelativistic limit $|\vec{k}|\gg m_{1},m_{2},m_{3}$.
Then, the oscillation formulae are highly spin-dependent,  $\mathcal{Q^{\uparrow}}_{\nu_{\sigma}\rightarrow\nu_{\rho}}^{\vec{k}}(t)\neq
\mathcal{Q^{\downarrow}}_{\nu_{\sigma}\rightarrow\nu_{\rho}}^{\vec{k}}(t)$, since in QFT framework, the oscillation amplitudes and the frequencies
are spin depending. Notice that, in QM mixing treatment, the spin orientation affects only the frequencies $\Delta_{ij}$, being in this case:
$\Gamma_{ij;\vec{k}}^{\pm \pm} = 1$, $\Sigma_{ij;\vec{k}}^{\pm \pm} =  \Omega_{ij;\vec{k}}^{\pm} = 0 $.

In the following, we analyze the behaviour of the oscillation formulae for constant and for time dependent torsions.

\subsection{Neutrino oscillation with constant torsion}

We report the transition formulas for sample values of torsion and momentum. We consider values of neutrino
masses $m_{1}\approx10^{-3}\mathrm{eV}$, $m_{2}\approx9\times10^{-3}\mathrm{eV}$,
and $m_{3}\approx2\times10^{-2}\mathrm{eV}$, in order that $\Delta m_{12}^{2}\approx7.56\times10^{-5}\mathrm{eV}^{2}$
and $\Delta m_{23}^{2}\approx2.5\times10^{-3}\mathrm{eV}^{2}$, and
of mixing angles such that $\sin^{2}(2\theta_{13})=0.10,$ $\sin^{2}(2\theta_{23})=0.97,$
and $\sin^{2}(2\theta_{12})=0.861$, which are compatible with the experimental data. We also consider $\delta=\pi/4$,
and a  fixed value of the momentum $k\simeq2\times10^{-2}\mathrm{eV}$
and of the torsion $|T^{3}|\simeq2\times10^{-4}\mathrm{eV}$.
In figs.1,2 and 3, we plot  $\mathcal{Q^{\uparrow}}_{\nu_{e}\rightarrow\nu_{\sigma}}^{\vec{k}}(t)$
  and $\mathcal{Q^{\downarrow}}_{\nu_{e}\rightarrow\nu_{\sigma}}^{\vec{k}}(t)$, with $\sigma = e, \mu, \tau$,
  as a function of time, and we compare such formulae with the corresponding quantum mechanics ones.
Such formulae can be obtained from   eqs.$(\ref{eq: Formula di oscillazione 3 generazioni Pe-e})$,
$(\ref{eq: Formula di oscillazione 3 generazioni Pe-mu})$, $(\ref{eq: Formula di oscillazione 3 generazioni Pe-tau})$, by setting
$\Gamma_{ij;\vec{k}}^{\pm \pm} = 1$, $\Sigma_{ij;\vec{k}}^{\pm \pm} = 0$ and  $\Omega_{ij;\vec{k}}^{\pm} = 0 $.

\begin{figure}[H]
\begin{centering}
\includegraphics[scale=0.7]{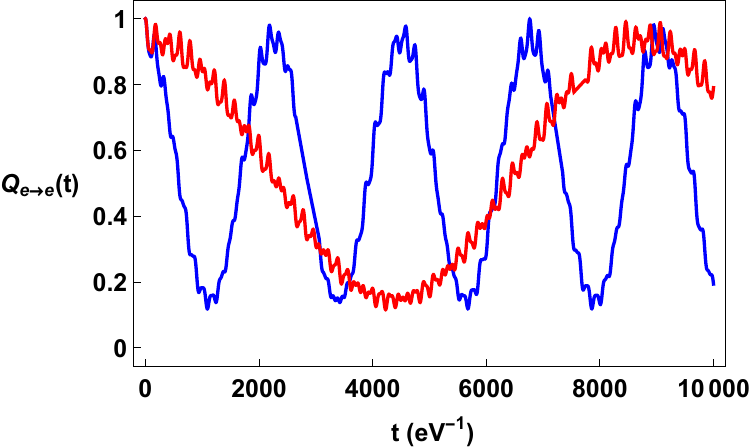}\includegraphics[scale=0.7]{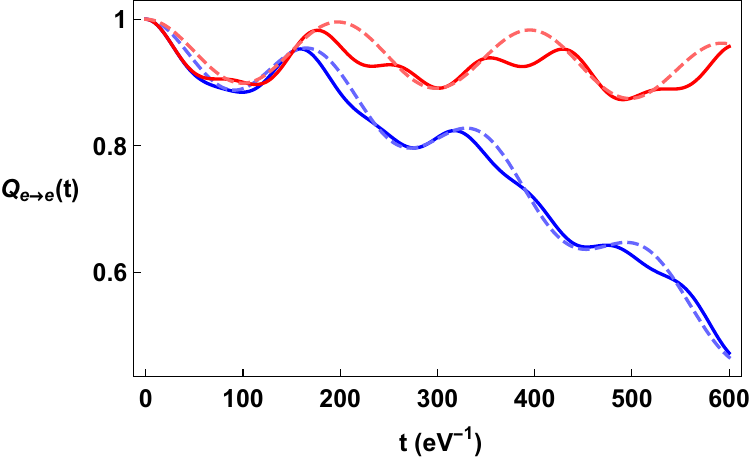}
\par\end{centering}
\caption{\label{fig:Pe-e. z50. 3 generazioni} Color on line. Plots of the oscillation formulae
in a constant torsion background: in the left-hand panel $\mathcal{Q^{\uparrow}}_{\nu_{e}\rightarrow\nu_{e}}^{\vec{k}}(t)$
(blue line) and $\mathcal{Q^{\downarrow}}_{\nu_{e}\rightarrow\nu_{e}}^{\vec{k}}(t)$
(red line) as a function of time. Torsion was picked to be comparable to
the momentum as ${T}^{3}=2\times10^{-4}\ \mathrm{eV}$. In the right
panel is reported the detail of the same formulae and the  comparison with the corresponding
quantum mechanics oscillation formulae (dashed line).}
\end{figure}
\begin{figure}[H]
\begin{centering}
\includegraphics[scale=0.7]{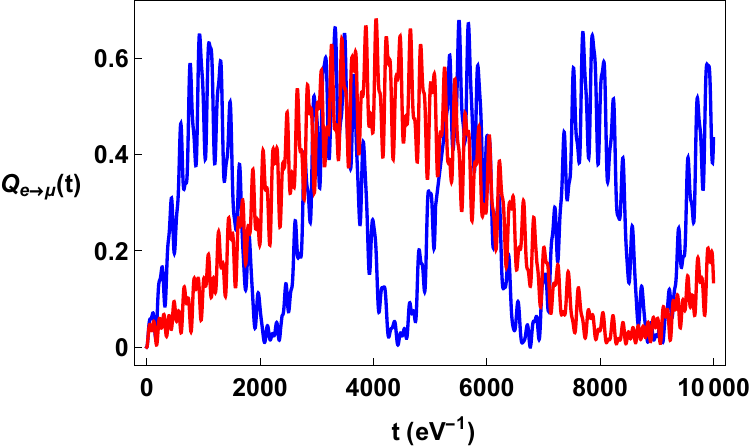}\includegraphics[scale=0.7]{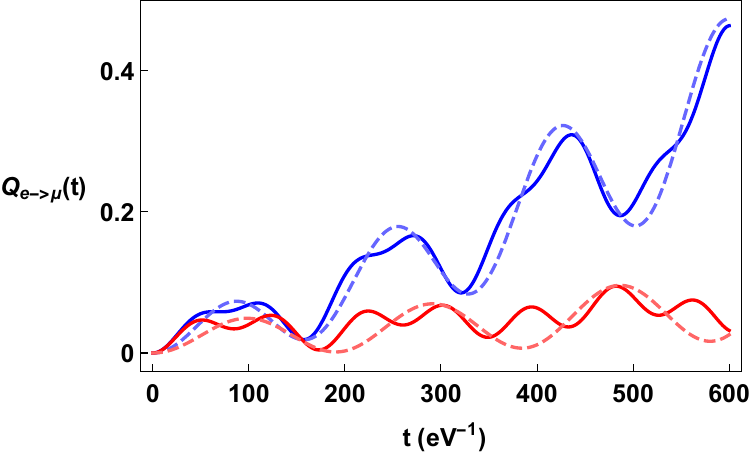}
\par\end{centering}
\caption{\label{fig::1} Color on line. In the left-hand panel, plot of $\mathcal{Q^{\uparrow}}_{\nu_{e}\rightarrow\nu_{\mu}}^{\vec{k}}(t)$
(blue line) and $\mathcal{Q^{\downarrow}}_{\nu_{e}\rightarrow\nu_{\mu}}^{\vec{k}}(t)$
(red line) as a function of time.  In the right
panel, detail of the same formulae and comparison with the corresponding QM
oscillation formulae (dashed line).}
\end{figure}
 \begin{figure}[H]
\begin{centering}
\includegraphics[scale=0.7]{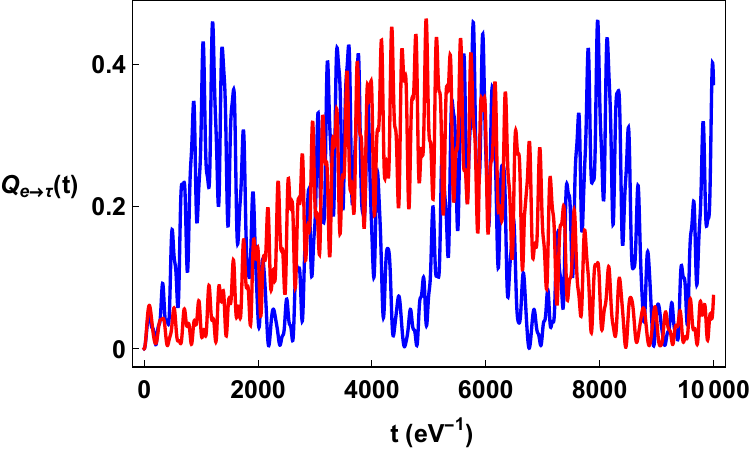}\includegraphics[scale=0.7]{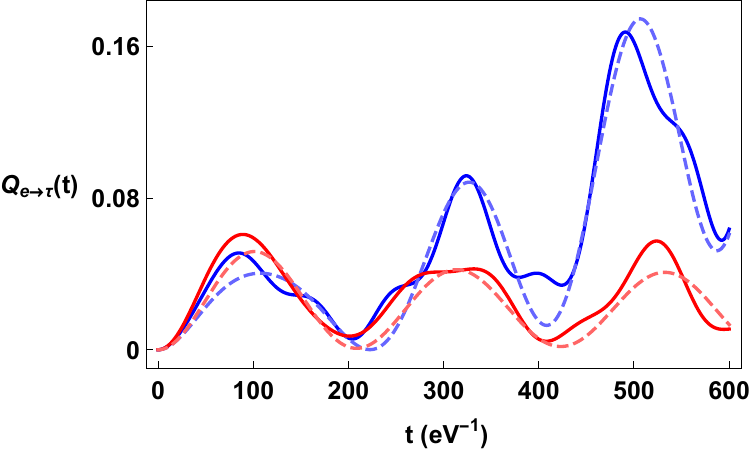}
\par\end{centering}
\caption{\label{fig:Pe-mu. 3 generazioni. z50} Color on line.  In the left-hand panel, plot of   $\mathcal{Q^{\uparrow}}_{\nu_{e}\rightarrow\nu_{\tau}}^{\vec{k}}(t)$
(blue line) and $\mathcal{Q^{\downarrow}}_{\nu_{e}\rightarrow\nu_{\tau}}^{\vec{k}}(t)$
(red line) as a function of time.  In the right
panel, detail of the same formulae and comparison with the corresponding QM
 formulae (dashed line).}
\end{figure}
The plots of the neutrino oscillation formulae for the constant torsion background displayed in figs.(1), (2) and (3) show the strong  dependence of them   on the spin orientation.   The difference  is maximal when the torsion is
comparable with the neutrino momentum and neutrino masses.
On the other hand, for very big values
of torsion, $T^{3}\gg m_{i},|\vec{k}|$, the energy terms are dominated by the torsion, indeed
$(E_{\vec{k},i}^{\pm})^{2} = \vec{k}^{2}+(\widetilde{m}_{i}^{\pm})^{2}\simeq(\widetilde{m}_{i}^{\pm})^{2}\simeq\left(\pm\frac{3}{2}T^{3}\right)^{2}$,
so that $E^{+}\simeq E^{-}$. This implies that both the Bogoliubov coefficients
$\Xi^{rr},\,\chi^{rr}$ and the phase factors $\Delta^{r},\,\Omega^{r}$
become essentially independent of the spin, and the
 flavor oscillations become independent on the spin orientation.
 We also note that  a torsion large enough can effectively inhibit the flavor oscillations, since
for $T^{3}\gg m_{i}$, the energy differences due to the   mass  differences
(e.g. $\Delta m_{12}$, $\Delta m_{13}$ and $\Delta m_{23}$) become irrelevant with respect to the
common torsional energy term.

\subsection{Neutrino oscillations with time dependent torsion }

The neutrino oscillation formulae, in the case of linearly time-dependent
torsion for fixed momentum  $\vec{k}$ and spin ($\uparrow$), are given by Eqs.(\ref{eq: Formula di oscillazione 3 generazioni Pe-e},\ref{eq: Formula di oscillazione 3 generazioni Pe-mu},\ref{eq: Formula di oscillazione 3 generazioni Pe-tau})
with the Bogoliubov coefficients expressed in Eqs.$(\ref{eq: Bogoljubov U++ time-dependent})$,$(\ref{eq: Bogoljubov V++ time-dependent})$.
By utilizing the same values of the  masses, of the angles and of the momentum used in fig.(1), (2), and (3),
for constant torsion, we plot in fig. (4),  (5), and (6) the oscillation formulae for time dependent torsion. We assume
$\eta\breve{T}^{0} \simeq 5 \times 10^{-3}\mathrm{eV}$. We observe that, also in this case, the formulas strongly depend on the orientation of the spin. In the computations here presented, we neglected the spin-flip transition
due to the torsion term. This analysis will be carried out in a forthcoming
work.

\begin{figure}[H]
\begin{centering}
\includegraphics[scale=0.7]{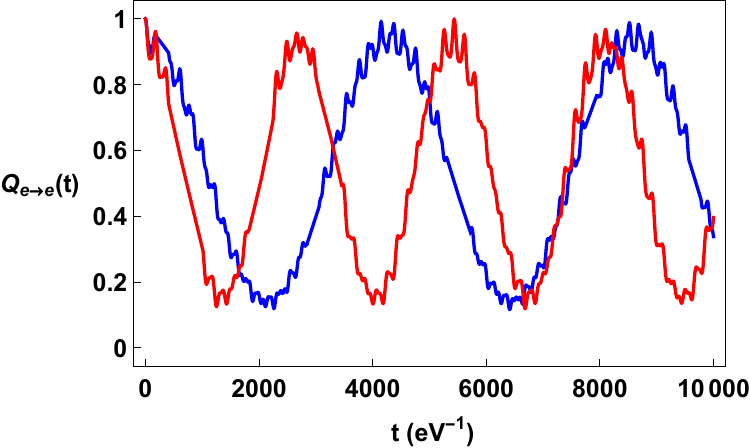}\includegraphics[scale=0.7]{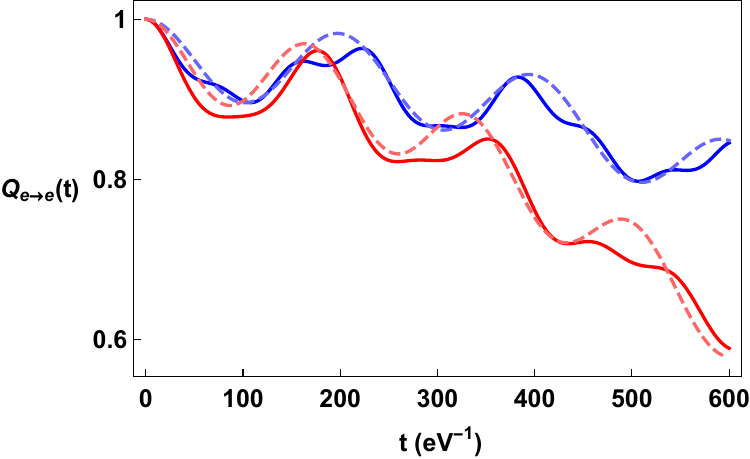}
\par\end{centering}
\caption{\label{fig::time torsione 0,1} Color on line. Plots of the oscillation
formulae in a time-dependent torsion: in the left-hand panel are plotted $\mathcal{Q^{\uparrow}}_{\nu_{e}\rightarrow\nu_{e}}^{\vec{k}}(t)$
(blue line) and $\mathcal{Q^{\downarrow}}_{\nu_{e}\rightarrow\nu_{e}}^{\vec{k}}(t)$
(red line) as a function of time. In the right
panel is reported the detail of the same formulae and the  comparison with the corresponding
quantum mechanics oscillation formulae (dashed line). We consider
 $\eta\breve{T}^{0}=5 \times 10^{-3}\ \mathrm{eV}$. }
\end{figure}
\begin{figure}[H]
\begin{centering}
\includegraphics[scale=0.7]{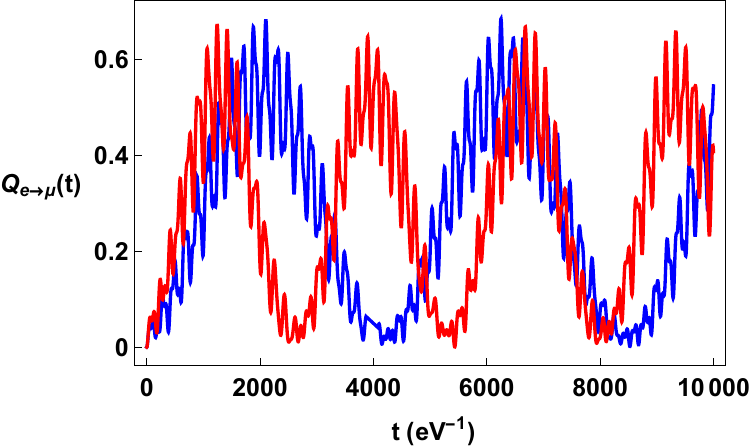}\includegraphics[scale=0.7]{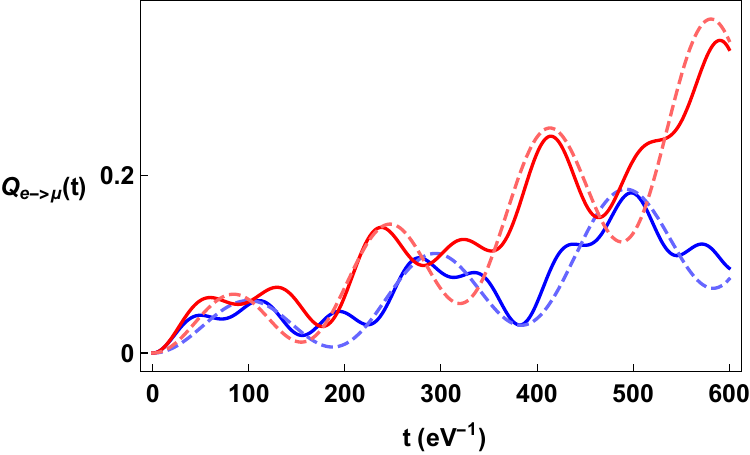}\caption{{\label{fig: probabilit=0000E0 torsione 0,01} Color online. In the left-hand panel plot  of $\mathcal{Q^{\uparrow}}_{\nu_{e}\rightarrow\nu_{\mu}}^{\vec{k}}(t)$
(blue line) and $\mathcal{Q^{\downarrow}}_{\nu_{e}\rightarrow\nu_{\mu}}^{\vec{k}}(t)$
(red line) as a function of time.   In the right
panel is reported the detail of the same formulae and the  comparison with the corresponding
QM   formulae (dashed line). }}
\par\end{centering}
\end{figure}
\begin{figure}[H]
\begin{centering}
\includegraphics[scale=0.7]{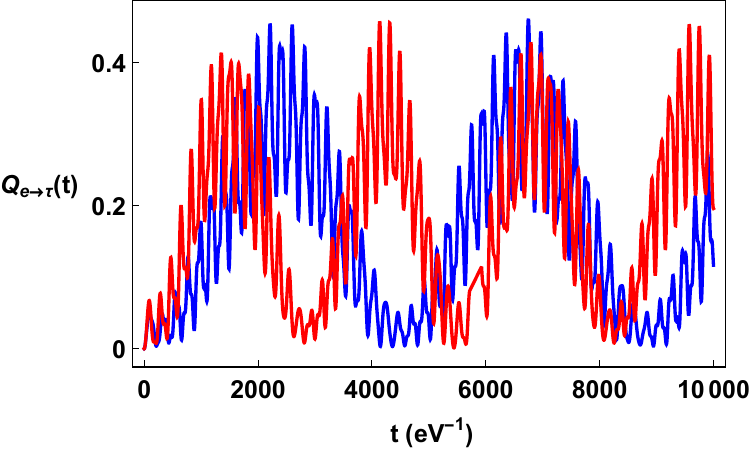}\includegraphics[scale=0.7]{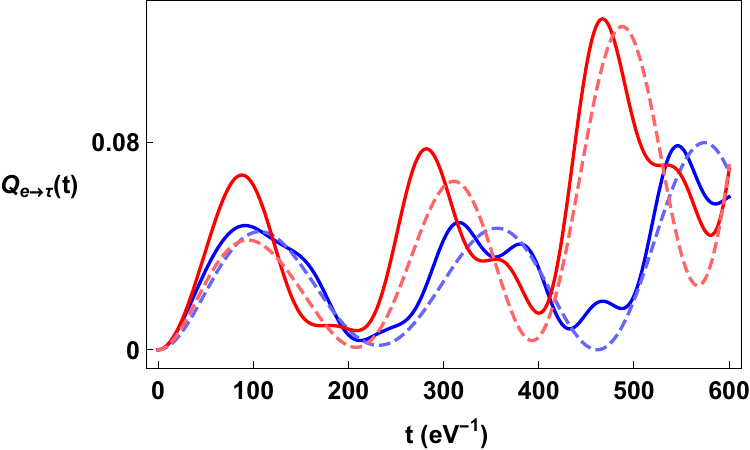}
\par\end{centering}
\caption{{\label{fig:petautime3sapori}Color online. In the left-hand panel plot of $\mathcal{Q^{\uparrow}}_{\nu_{e}\rightarrow\nu_{\tau}}^{\vec{k}}(t)$
(blue line) and $\mathcal{Q^{\downarrow}}_{\nu_{e}\rightarrow\nu_{\tau}}^{\vec{k}}(t)$
(red line) as a function of time.    In the right
panel is reported the detail of the same formulae and the  comparison with the corresponding
QM   formulae (dashed line).}}
\end{figure}

\section{$CP$ Violation and flavor vacuum}

We now  study the impact of torsion on the $CP$ violation in neutrino oscillation due to the presence of Dirac phase
in the mixing matrix. For fixed spin orientation, say $\uparrow$,  the $CP$ asymmetry $\Delta_{\uparrow;CP}^{\rho\sigma}$ can be defined in QFT,
in a similar way to what was done in the ref. \cite{N5}, and then:
$
\Delta_{\uparrow;CP}^{\rho\sigma}(t)  \equiv\mathcal{Q^{\uparrow}}_{\nu_{\rho}\rightarrow\nu_{\sigma}}^{\vec{k}}(t)+
\mathcal{Q^{\uparrow}}_{\overline{\nu}_{\rho}\rightarrow\overline{\nu}_{\sigma}}^{\vec{k}}(t)\,,
$
where $\rho,\sigma=e,\mu,\tau$.  Notice that
a $+$ sign appears in front of the probabilities for the antineutrinos,
in place of   $-$, because the antineutrino states already
carry a negative flavor charge $Q_{\sigma}$.
%
For the $\nu_{e}\rightarrow\nu_{\mu}$ transition, with $r=\uparrow,\downarrow$, the $CP$ asymmetry is explicitly
\begin{align}
\Delta_{r;CP}^{e\mu}(t) & =4J_{CP}\left[\left|\Gamma_{12;\vec{k}}^{\pm \pm}\right|^{2}\sin\left(2\Delta_{12;\vec{k}}^{\pm}t\right)-
\left|\Sigma_{12;\vec{k}}^{\pm \pm}\right|^{2}\sin\left(2\Omega_{12;\vec{k}}^{\pm}t\right)+\right.
+\left(\left|\Gamma_{12;\vec{k}}^{\pm \pm}\right|^{2}-
 \left|\Sigma_{13;\vec{k}}^{\pm \pm}\right|^{2}\right)\sin\left(2\Delta_{23;\vec{k}}^{\pm}t\right)\nonumber \\
 &+
 \left(\left|\Sigma_{12;\vec{k}}^{\pm \pm}\right|^{2}-\left|\Sigma_{13;\vec{k}}^{\pm \pm}\right|^{2}\right)\sin\left(2\Omega_{23;\vec{k}}^{\pm}t\right)
  \left.-\left|\Gamma_{13;\vec{k}}^{\pm \pm}\right|^{2}\sin\left(2\Delta_{13;\vec{k}}^{\pm}t\right)+
 \left|\Sigma_{13;\vec{k}}^{\pm \pm}\right|^{2}\sin\left(2\Omega_{13;\vec{k}}^{\pm}t\right)\right]\,,\label{eq: CPup}
\end{align}
where one has to consider  $\Gamma_{ij;\vec{k}}^{++}$ and $ \Sigma_{ij;\vec{k}}^{++}$ for spin up and
$\Gamma_{ij;\vec{k}}^{--}$ and $ \Sigma_{ij;\vec{k}}^{--}$ for spin down.
One also has $\Delta_{r;CP}^{e\tau}(t)=-\Delta_{r;CP}^{e\mu}(t)$ with
$r=\uparrow,\downarrow$.
Remarkably the presence of torsion induces a  $CP$ asymmetry depending on the spin orientation.

Furthermore, we  make some observations on the condensate structure of the flavor vacuum in the presence of torsion. In this case,
 $\ket{0_{f}(t)}$   breaks the spin symmetry, resulting in a different
condensation density for particles of spin up and down.
Such  densities are evaluated by computing the expectation
values of the number operators for free fields
$N_{\alpha_{j},\vec{k}}^{r} = \alpha_{\vec{k},j}^{r\dagger}\alpha_{\vec{k},j}^{r}$
and $N_{\beta_{j},\vec{k}}^{r}=\beta_{\vec{k},j}^{r\dagger}\beta_{\vec{k},j}^{r}$,
on $\ket{0_{f}(t)}$.
One has
\begin{align}
\mathcal{N}_{1;\vec{k}}^{r} & =_{f}\left\langle 0(t)\right|\text{\ensuremath{N_{\alpha_{1},\vec{k}}^{r}}}\left|0(t)\right\rangle _{f}={}_{f}\left\langle 0(t)\right|\text{\ensuremath{N_{\beta_{1},\vec{k}}^{r}}}\left|0(t)\right\rangle _{f}  = s_{12}^{2}c_{13}^{2}\left|\Sigma_{12;\vec{k}}^{\pm \pm}\right|^{2}+s_{13}^{2}\left|\Sigma_{13;\vec{k}}^{\pm \pm}\right|^{2}\,,\label{eq:condensato m1}
\\
\mathcal{N}_{2;\vec{k}}^{r} & =_{f}\left\langle 0(t)\right|\text{\ensuremath{N_{\alpha_{2},\vec{k}}^{r}}}\left|0(t)\right\rangle _{f}={}_{f}\left\langle 0(t)\right|\text{\ensuremath{N_{\beta_{2},\vec{k}}^{r}}}\left|0(t)\right\rangle _{f}
 =\left|-s_{12}c_{23}+e^{i\delta}c_{12}s_{23}s_{13}\right|^{2}\left|\Sigma_{12;\vec{k}}^{\pm \pm}\right|^{2}+s_{23}^{2}c_{13}^{2}
 \left|\Sigma_{23;\vec{k}}^{\pm \pm}\right|^{2}\,,\label{eq:condensato m2}
\\ \nonumber
\mathcal{N}_{3;\vec{k}}^{r} & =_{f}\left\langle 0(t)\right|\text{\ensuremath{N_{\alpha_{3},\vec{k}}^{r}}}\left|0(t)\right\rangle _{f}={}_{f}\left\langle 0(t)\right|\text{\ensuremath{N_{\beta_{3},\vec{k}}^{r}}}\left|0(t)\right\rangle _{f} \\
 &=\left|-c_{12}s_{23}+e^{i\delta}s_{12}c_{23}s_{13}\right|^{2}\left|\Sigma_{23;\vec{k}}^{\pm \pm}\right|^{2}+
 \left|s_{12}s_{23}+e^{i\delta}c_{12}c_{23}s_{13}\right|^{2}\left|\Sigma_{13;\vec{k}}^{\pm \pm}\right|^{2}\,\ ,\label{eq: condensato m3}
\end{align}
where,
$r=\uparrow,\downarrow$.

\subsection{$CP$ violation and flavor vacuum condensate with constant torsion}

For constant torsion, we plot in fig.$(7)$,  $\Delta_{\uparrow;CP}^{e\mu}(t)$ and $\Delta_{\downarrow;CP}^{e\mu}(t)$
as a function of time and in fig.$(8)$ we plot $\mathcal{N}_{i;\vec{k}}^{\uparrow}$
and $\mathcal{N}_{i;\vec{k}}^{\downarrow}$ with $i=1,2,3$, as a function of $\left|\vec{k}\right|$.
We use  the same values of the parameters adopted
in the plots of the oscillation formulae.
\begin{figure}[h]
\centering{}\includegraphics[scale=0.8]{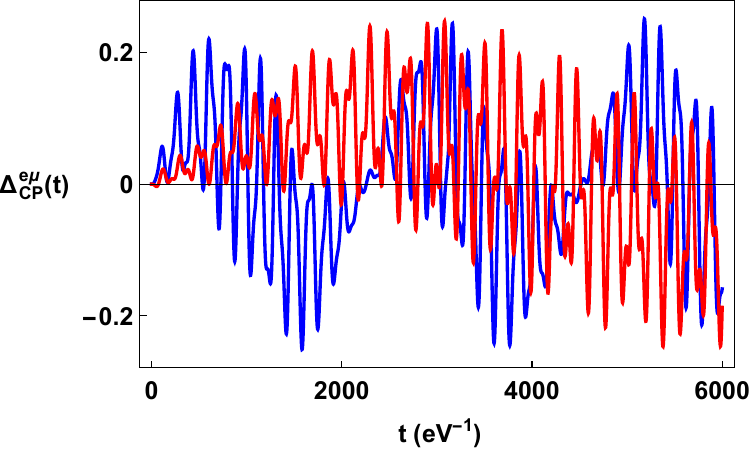}\caption{\label{fig:Asymmetry. z50} Color on line. Plot of $\Delta_{\uparrow;CP}^{e\mu}(t)$
(blue line) and $\Delta_{\downarrow;CP}^{e\mu}(t)$ (red line) as a function of
time for the values of the  parameters used in Figs. (1), (2) and (3).}
\end{figure}
\begin{figure}[h]
\begin{centering}
\includegraphics[scale=0.6]{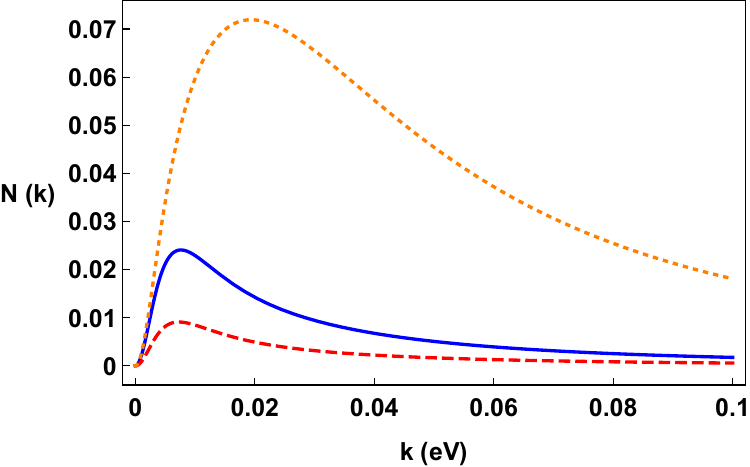}\includegraphics[scale=0.6]{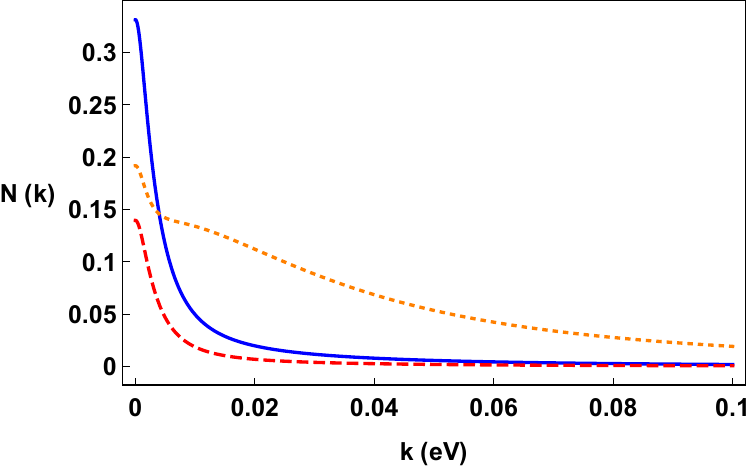}
\par\end{centering}
\caption{\label{fig:z50condensato} Color online.  (Left panel) Plots of $\mathcal{N}_{i;\vec{k}}^{\uparrow}$
as a function of $\left|\vec{k}\right|$ in $\mathrm{eV}$: $N_{1}$
(Blue solid), $N_{2}$ (Red dashed line) and $N_{3}$ (Orange dotted line). (Right panel) Plots of $\mathcal{N}_{i;\vec{k}}^{\downarrow}$
as a function of $\left|\vec{k}\right|$. We use the same  parameters adopted in Figs. (1), (2) and (3).}
\end{figure}

\subsection{$CP$ violation and flavor vacuum condensate for time dependent torsion}

In the case of linearly time-dependent torsion, the  $CP$ violation and the condensation densities are plotted in figs.(9) and (10), respectively,
for the same values of the parameters used for figs. (4), (5) and (6).

\begin{figure}[H]
\begin{centering}
\includegraphics[bb=0bp 0bp 475bp 265bp,scale=0.8]{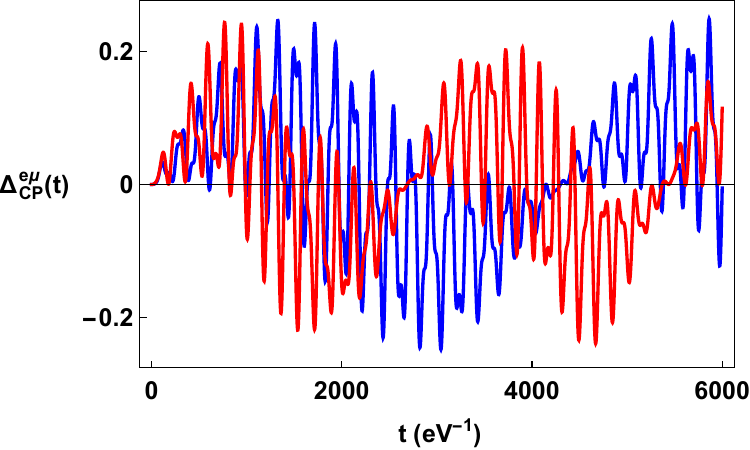}
\par\end{centering}
\caption{{\label{fig:asimmetriatime}Color on line. Plot of $\Delta_{\uparrow;CP}^{e\mu}(t)$
(blue) $\Delta_{\downarrow;CP}^{e\mu}(t)$ (red) as a function of
time for the same parameters used in figs. (4), (5) and (6).}}
\end{figure}
\begin{figure}[H]
\begin{centering}
\includegraphics[scale=0.7]{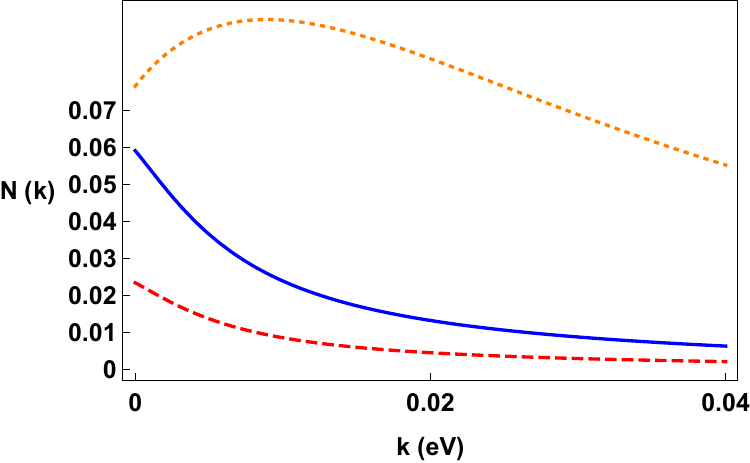}\includegraphics[scale=0.7]{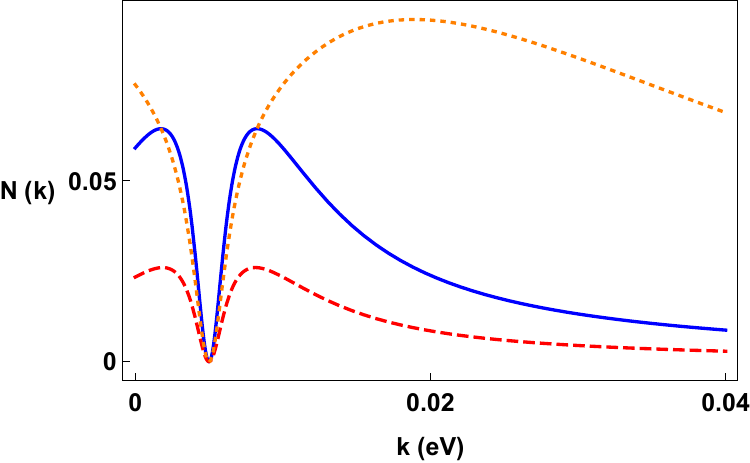}\caption{{\label{fig: condensato time}Color online. In the left panel, we plots   $\mathcal{N}_{i;\vec{k}}^{\uparrow}$
as a function of $\left|\vec{k}\right|$: $N_{1}$
in blue line, $N_{2}$  in red line and $N_{3}$ orange line, for the same values
of the parameters used in figs. (4), (5) and (6). In the right
panel) Plots of $\mathcal{N}_{i;\vec{k}}^{\downarrow}$ as a function
of $\left|\vec{k}\right|$ for the same choice of parameters.}}
\par\end{centering}
\selectlanguage{english}%
\selectlanguage{italian}%
\end{figure}
It is worth noting that the well shape appearing in the right panel of Fig. (10) is due to the proportionality of $\varUpsilon^{--}_{i,j, \vec{p}}$ to $(p-\eta \breve{T}^0)$ (see Eq. \eqref{eq: Bogoljubov V++ time-dependent}), so that it vanishes for $p= \eta \breve{T}^0$, bringing also the condensation density to zero. 
\selectlanguage{english}%

\section{Conclusions}

We analyzed the Einstein-Cartan theory and by studying the neutrino
propagation on a torsion background in the framework QFT, we derived
new oscillation formulae which are depending on the spin orientations
of the neutrino fields. Indeed, we have shown that the energy splitting
induced by the torsion term affects the oscillation frequencies and
the Bogoliubov coefficients which represent the amplitudes of the
oscillation formulae. We considered flat space-time and two different kind of
of torsion terms, the constant and the linearly time dependent torsion.

The two analyzed cases share the following behavior: the  spin dependence of the oscillation is maximal for values of torsion
comparable to the neutrino momentum and masses, while much larger
values of torsion lead to flavor oscillations which are almost independent
of the spin. Moreover, a torsion large enough can effectively inhibit
the flavor oscillations. Such behaviours characterize also the $CP$-asymmetry.

The torsion effects are relevant on neutrino oscillations in non-relativistic
regimes. Therefore, experiments studying neutrinos with very low momenta,
such as PTOLEMY, could provide verification of such results in the
future.

\section{Appendix A: Useful formulae }

For reader convenience, we report formulas  useful for the computations. We consider the PNMS matrix matrix. Then, denoting with $\psi_{f}^{T}=\left(\nu_{e},\nu_{\mu},\nu_{\tau}\right)$, the flavor fields
and with $\Psi_{m}^{T}=\left(\nu_{1},\nu_{2},\nu_{3}\right)$ the fields with definite masses, the mixing relations are:
\[
\Psi_{f}(x)=\left(\begin{array}{ccc}
c_{12}c_{13} & s_{12}c_{13} & s_{13}e^{-i\delta}\\
-s_{12}c_{23}-c_{12}s_{23}s_{13}e^{i\delta} & c_{12}c_{23}-s_{12}s_{23}s_{13}e^{i\delta} & s_{23}c_{13}\\
s_{12}s_{23}-c_{12}c_{23}s_{13}e^{i\delta} & -c_{12}s_{23}-s_{12}c_{23}s_{13}e^{i\delta} & c_{23}c_{13}
\end{array}\right)\Psi_{m}(x)\ .
\]
Here,  $c_{ij}=\cos\theta_{ij}, s_{ij}=\sin\theta_{ij}$
 and $\delta$ is the Dirac $CP$-violating phase.

 The mixing generator $\emph{I}_{\theta}$ is given by $\emph{I}_{\theta}(t)=\emph{I}_{23}(t)\emph{I}_{13}(t)\emph{I}_{12}(t)\;,$
where
\[
\begin{array}{c}
\emph{I}_{12}(t)\equiv\exp\left[\theta_{12}\int d^{3}\boldsymbol{x}\left(\nu_{1}^{\dagger}(x)\nu_{2}(x)-\nu_{2}^{\dagger}(x)\nu_{1}(x)\right)\right]\,,\\
\emph{I}_{23}(t)\equiv\exp\left[\theta_{23}\int d^{3}\boldsymbol{x}\left(\nu_{2}^{\dagger}(x)\nu_{3}(x)-\nu_{3}^{\dagger}(x)\nu_{2}(x)\right)\right]\,,\\
\emph{I}_{13}(t)\equiv\exp\left[\theta_{13}\int d^{3}\boldsymbol{x}\left(\nu_{1}^{\dagger}(x)\nu_{3}(x)e^{-i\delta}-\nu_{3}^{\dagger}(x)\nu_{1}(x)e^{i\delta}\right)\right]\,,
\end{array}
\]
with $\nu_{i}$ free fields solutions of Dirac equations with torsion terms.

 The Bogoliubov coefficients $\Gamma_{ij;\vec{k}}^{rs}$ and $\Sigma_{ij;\vec{k}}^{rs}$
  satisfy the following identities:
\[
\begin{array}{c}
\Sigma_{23;\vec{k}}^{rr}(t)\left(\Sigma_{13;\vec{k}}^{rr}(t)\right)^{*}+\left(\Gamma_{23;\vec{k}}^{rr}(t)\right)^{*}\Gamma_{13;\vec{k}}^{rr}(t)
=\Gamma_{12;\vec{k}}^{rr}(t)\,,\,\,\,\,\,\,\,
\Sigma_{23;\vec{k}}^{rr}(t)\left(\Gamma_{13;\vec{k}}^{rr}(t)\right)^{*}-\left(\Gamma_{23;\vec{k}}^{rr}(t)\right)^{*}\Sigma_{13;\vec{k}}^{rr}(t)
=-\Sigma_{12;\vec{k}}^{rr}(t)\,,
\\
\Gamma_{12;\vec{k}}^{rr}(t)\Gamma_{23;\vec{k}}^{rr}(t)-\left(\Sigma_{12;\vec{k}}^{rr}(t)\right)^{*}\Sigma_{23;\vec{k}}^{rr}(t)
=
\Gamma_{13;\vec{k}}^{rr}(t)\,,\,\,\,\,\,\,\,
\Gamma_{23;\vec{k}}^{rr}(t)\Sigma_{12;\vec{k}}^{rr}(t)+\left(\Gamma_{12;\vec{k}}^{rr}(t)\right)^{*}\Sigma_{23;\vec{k}}^{rr}(t)=
\Sigma_{13;\vec{k}}^{rr}(t)\,,
\\
\left(\Sigma_{12;\vec{k}}^{rr}(t)\right)^{*}\Sigma_{13;\vec{k}}^{rr}(t)+\left(\Gamma_{12;\vec{k}}^{rr}(t)\right)^{*}\Gamma_{13;\vec{k}}^{rr}(t)
=
\Sigma_{23;\vec{k}}^{rr}(t)\,,\,\,\,\,\,\,\,
\Sigma_{12;\vec{k}}^{rr}(t)\Sigma_{13;\vec{k}}^{rr}(t)-\Sigma_{12;\vec{k}}^{rr}(t)\Sigma_{13;\vec{k}}^{rr}(t)=-\Sigma_{12;\vec{k}}^{rr}(t)\,,
\\
\xi_{13;\vec{k}}^{rr}=\xi_{12;\vec{k}}^{rr}+\xi_{23;\vec{k}}^{rr},\quad \,\,\, \xi_{ij;\vec{k}}^{rr}=
\arctan\left(\frac{\left|\Sigma_{ij;\vec{k}}^{rr}\right|}{\left|\Gamma_{ij;\vec{k}}^{rr}\right|}\right)\,.
\end{array}
\]
\\

\section*{Appendix B: CHARGES FOR THREE FLAVOR MIXING WITH TORSION}

Charges are introduced, by using the symmetries of the Lagrangian for free field operators: $L = \overline{\psi}_{m}(x)(i \gamma_\mu \partial^\mu - M){\psi}_{m}(x)$. The Lagrangian is invariant under global
transformation of a phase factor $U(1)$ of the type $\Psi_{m}'=e^{i\alpha}\Psi_{m}$. Then
a charge is introduced via Noether's theorem: $Q=\int d^{3}\boldsymbol{x}\overline{\Psi}_{m}(x)\gamma^{0}\Psi_{m}$;
it represents the total charge of the system. Considering a field
transformation $\Psi_{m}$ under global transformation $SU(3)$, we
obtain Noether charges $Q_{m,j}$ of the form: $Q_{m,j}(t)\equiv\int d^{3}\boldsymbol{x}J_{m,j}^{0}(x)\,,$
with $j=1,2,\cdots,8$ and $J_{m,j}^{0}(x)$, time component of the $SU(3)$ currents. The charges satisfy the $SU(3)$ algebra:
$\left[Q_{m,j}(t),Q_{m,k}(t)\right]=if_{jkl}Q_{m,l}(t)\,.$ Note that
only charges $Q_{m,3}$ and $Q_{m,8}$ are not time-dependent. Appropriate
combinations of these charges allow to define the  quantities: $Q_{1}\equiv\frac{1}{3}Q+Q_{m,3}+\frac{1}{\sqrt{3}}Q_{m,8}\;,$
$Q_{2}\equiv\frac{1}{3}Q-Q_{m,3}+\frac{1}{\sqrt{3}}Q_{m,8}\;,$ and
$Q_{3}\equiv\frac{1}{3}Q-\frac{2}{\sqrt{3}}Q_{m,8}\,.$ 
The normal ordering of charge operators for free fields are then:
$
\left.:\,Q_{i}\,:\right.\equiv\sum_{r}\int d^{3}\boldsymbol{k}\,\left(\alpha_{\vec{k},i}^{r\dagger}\alpha_{\vec{k},i}^{r}-\beta_{-\vec{k},i}^{r\dagger}\beta_{-\vec{k},i}^{r}\right)\,$
with $ i=1,2,3
$
 where $\left.:\cdots:\right.$ has been used to denote the normal
ordered with respect to the vacuum state $\left|0\right\rangle _{m}$.

The flavour charges can be directly derived from    the above Noether charges by applying the mixing generator to them: $::\,Q_{\nu_{\sigma}}\,(t)::=\emph{I}_{\theta}^{-1}(t):\,Q_{i}\,:\emph{I}_{\theta}(t)\;,$
with $(\sigma,i)=(e,1),(\mu,2),(\tau,3)$.
%
In terms
of the flavour annihilators one has:
\[
::\,Q_{\nu_{\sigma}}\,::=\sum_{r}\int d^{3}\boldsymbol{k}\left(\alpha_{\vec{k},\nu_{\sigma}}^{r\dagger}(t)\alpha_{\vec{k},\nu_{\sigma}}^{r}(t)-\beta_{\vec{k},\nu_{\sigma}}^{r\dagger}(t)\beta_{\vec{k},\nu_{\sigma}}^{r}(t)\right)\,,\qquad\sigma=e,\mu,\tau
\]
\\
 where $\left.::\cdots::\right.$ the normal ordered with respect
to the vacuum state was indicated $\left|0\right\rangle _{f}$. \\
 \\
The neutrino
oscillation formula at a fixed momentum $\vec{k}$ and spin $(\uparrow)$ are obtained in the Heisenberg picture, by computing the following expectation values:
\begin{align*}
\mathcal{Q^{\uparrow}}_{\nu_{\rho}\rightarrow\nu_{\sigma}}^{\vec{k}}(t) & \equiv\left\langle \nu_{\vec{k},\rho}^{\uparrow}(t)\right|::\,Q_{\nu_{\sigma}}\,::\left|\nu_{\vec{k},\rho}^{\uparrow}(t)\right\rangle -{}_{f}\left\langle 0\right|::\,Q_{\nu_{\sigma}}\,::\left|0\right\rangle _{f}\\
 & =\left|\left\{ \alpha_{\vec{k},\nu_{\sigma}}^{\uparrow}(t),\alpha_{\vec{k},\nu_{\rho}}^{\uparrow\dagger}(0)\right\} \right|^{2}+\left|\left\{ \beta_{-\vec{k},\nu_{\sigma}}^{\uparrow\dagger}(t),\alpha_{\vec{k},\nu_{\rho}}^{\uparrow\dagger}(0)\right\} \right|^{2}\,.
\end{align*}
Similarly for the antiparticle:
\begin{align*}
\mathcal{Q^{\uparrow}}_{\overline{\nu}_{\rho}\rightarrow\overline{\nu}_{\sigma}}^{\vec{k}}(t) & \equiv\left\langle \overline{\nu}_{\vec{k},\rho}^{\uparrow}(t)\right|::\,Q_{\nu_{\sigma}}\,::\left|\overline{\nu}_{\vec{k},\rho}^{\uparrow}(t)\right\rangle -{}_{f}\left\langle 0\right|::\,Q_{\nu_{\sigma}}\,::\left|0\right\rangle _{f}\\
 & =-\left|\left\{ \beta_{\vec{k},\nu_{\sigma}}^{\uparrow}(t),\beta_{\vec{k},\nu_{\rho}}^{\uparrow\dagger}(0)\right\} \right|^{2}-\left|\left\{ \alpha_{-\vec{k},\nu_{\sigma}}^{\uparrow\dagger}(t),\beta_{\vec{k},\nu_{\rho}}^{\uparrow\dagger}(0)\right\} \right|^{2}\,.
\end{align*}
Similar formulae are obtained for spin down.

\section*{Acknowledgements}

Partial financial support from MUR and INFN is acknowledged. A.C.
also acknowledges the COST Action CA1511 Cosmology and Astrophysics
Network for Theoretical Advances and Training Actions (CANTATA).


\end{document}